\documentclass[%
 reprint,
 amsmath,amssymb,
 aps,
 pra,
 floatfix,
]{revtex4-2}

\usepackage{graphicx}%
\usepackage{dcolumn}%
\usepackage{bm}%
\usepackage[hidelinks]{hyperref}%
\usepackage{tikz}
\usetikzlibrary{arrows.meta, positioning, shapes.geometric, decorations.pathmorphing, decorations.pathreplacing, calc, fit, backgrounds}
\usepackage{natbib}
\usepackage{placeins}
\usepackage{etoolbox}
\AtBeginEnvironment{thebibliography}{\clubpenalty=10000\widowpenalty=10000}

\usepackage{mathtools}

\newtheorem{proposition}{Proposition}
\newtheorem{corollary}{Corollary}

\newcommand{\id}{I}
\newcommand{\proj}{\Pi}
\newcommand{\tr}{\operatorname{Tr}}
\newcommand{\rank}{\operatorname{rank}}

\newcommand{\diag}{\operatorname{diag}}

\newcommand{\projket}[1]{|#1\rangle\!\langle #1|}

\allowdisplaybreaks[2]

\begin{document}

	\title{Receiver-Constrained Recovery Placement for Encoded-Qutrit Teleportation under Monitored Decay}%

	\author{Yifan Shi}
	\author{Duan Wang}
	\author{Quanzhen Ding}
	\author{He Zhang}
	\author{Da-Wei Luo}
	\author{Ting Yu}
	\affiliation{Department of Physics and Center for Quantum Science and Engineering, Stevens Institute of Technology, Hoboken, New Jersey 07030, USA}
	\date{October 3, 2026}

\begin{abstract}
	
	Using established probabilistic measurement-reversal theory, we study recovery placement in encoded-qutrit teleportation under monitored decay. We characterize the success advantage of reversing a no-jump map only on the encoded support and compare two receiver capabilities. The restricted-map construction applies to general finite-dimensional supports; explicit timing results are obtained for a coherent $V$-type sender, a $\Lambda$-type receiver, and $S=\{0,2\}$, within a prescribed family of maximal exact sender reversals. With one outcome-independent full-space receiver filter, we maximize success probability at fixed total time subject to a minimum success-weighted conditional fidelity. A single preassigned recovery time yields up to three duration-dependent regimes: no sender reversal, the earliest feasible intermediate reversal, and final-time reversal. If recovery times may be randomized independently of the input and only the pooled successful ensemble must meet the fidelity requirement, the optimum over all such timing distributions mixes the two endpoints for $\Gamma T\geq(\ln3)/2$. With Bell-outcome-dependent nonunitary receiver recovery, each accepted full-column-rank transfer branch can instead be reversed exactly. For the same model and sender-operation family, omitting sender reversal uniquely maximizes the exact-teleportation success probability for every finite $T>0$ at positive decay rate. Partial pathway coherence further distinguishes relative support gain from absolute recovery success: the long-time dark-state plateau survives only for fully coherent decay. These results show how receiver control capabilities and the meaning of the fidelity constraint determine whether intermediate sender recovery is useful.

\end{abstract}

\maketitle

\section{Introduction}\label{sec.1}

Open quantum systems are inevitably influenced by environmental noise, and the resulting decoherence remains one of the main obstacles to reliable quantum information processing \cite{breuer2002theory,yu2004finite,yu2006quantum,wang2010sudden,bartkowiak2011sudden,you2005correlation}. In the broader setting of quantum networks and long-distance quantum communication \cite{briegel1998quantum,kimble2008quantum}, decoherence directly limits the faithful transfer, storage, and manipulation of quantum states. A variety of control and recovery strategies have therefore been developed, including feedback control \cite{ganesan2007decoherence,xue2012decoherence}, dynamical decoupling \cite{viola1999dynamical}, weak-measurement-based protection and reversal \cite{kofman2012nonperturbative,ashhab2010control,korotkov2010decoherence,singh2024experimental}, environment-assisted error correction \cite{trendelkamp2011environment}, and quantum error-correction protocols \cite{bennett1996mixed,ekert1996quantum,roffe2019quantum}. These approaches aim, in different ways, to suppress, correct, or probabilistically reverse the degradation induced by the environment.

Measurement-based protection includes distinct physical strategies. Continuous weak measurements of stabilizer operators, combined with Hamiltonian feedback, can protect an encoded state against errors \cite{ahn2002continuous}. More recently, history-dependent feedback has been used to adapt control operations to the measurement record in Gottesman--Kitaev--Preskill error correction \cite{puviani2025feedback}. A different strategy reverses a selected measurement branch probabilistically: Korotkov and Jordan proposed explicit solid-state implementations \cite{korotkov2006undoing}, and Katz \textit{et al.}\ demonstrated partial-collapse reversal in a superconducting phase qubit \cite{katz2008reversal}. Optical experiments further demonstrated amplitude-damping suppression by weak measurement before the noisy channel and reversal afterward, with improved recovery quality at the cost of reduced transmittance \cite{lee2011experimental}; related operations protect entanglement \cite{kim2012protecting}.

\begingroup\clubpenalty=10000\widowpenalty=10000
Environment-assisted error correction (EAEC) instead uses information extracted from the environment to guide system recovery \cite{gregoratti2003quantum,trendelkamp2011environment,memarzadeh2011recovering}. In the ideal random-unitary setting, suitable environment measurements and conditional unitary corrections permit exact restoration. For example, qubit and qutrit dephasing maps that preserve every diagonal observable in a fixed basis admit this recovery \cite{buscemi2005inverting}. General dissipative channels need not have a random-unitary decomposition. Environment-assisted weak measurement reversal (EA-WMR) addresses nonunitary branches by combining environmental monitoring with a probabilistic reversal \cite{wang2014environment}. In its standard full-space form, once a no-jump branch is selected, the recovery attempts to invert the corresponding Kraus operator on the entire carrier Hilbert space.
\par\endgroup

The singular-value limit on probabilistic reversal is established. For a known measurement branch, the maximum probability of exact reversal is controlled by its least singular value \cite{nielsen1997reversible,cheong2012balance}; probabilistic quantum error correction has also been formulated for general noise and encoding maps \cite{kukulski2023probabilistic}. More directly for teleportation, sender- and receiver-side weak measurements and reversals have been compared for amplitude-damped qubit resources \cite{qiu2014enhancing}; Ref.~\cite{lee2021teleportation} casts each teleportation outcome as an effective measurement followed by an outcome-dependent reversing operation, and Ref.~\cite{im2021optimal} optimizes and experimentally demonstrates such receiver-side reversal for noisy qubit channels. Feed-forward control, environment-assisted measurements, and reversal have also been combined to protect tripartite teleportation \cite{zhang2024tripartite}. In qubit entanglement swapping, Ref.~\cite{jeong2026recovery} further shows that reversal before the Bell measurement is unnecessary for selected $\Psi^\pm$ branches because the projection removes the population component targeted by that reversal.

\begingroup\clubpenalty=10000\widowpenalty=10000
Several related control questions have also been studied. Teleportation when both the unknown input and the resource channel are noisy has been optimized in a general framework \cite{taketani2012noisy}, and qutrit teleportation under correlated amplitude damping has been protected by weak- and environment-assisted measurements \cite{xiao2025enhanced}. The time of a single deterministic local NOT operation within amplitude-damping evolution has been used to steer entanglement decay and teleportation utility \cite{behera2026temporal,abhignan2026decoherence}, while Ref.~\cite{ali2026nearperfect} optimizes Alice's Bell-measurement time under non-Markovian dephasing and retains selected outcome classes. Interference between parallel decay paths of $V$-type qutrits has likewise been shown to protect teleportation fidelity \cite{anil2026interferences}. These works establish the recovery tools and control settings on which the present study builds. The question addressed below is narrower: for a fixed encoded qutrit support, fixed total time and qutrit Bell basis, and monitored local Markovian no-jump dynamics, how do the allowed receiver operation and the placement of one prescribed sender reversal change the optimal protocol? Unlike the deterministic temporal controls of Refs.~\cite{behera2026temporal,abhignan2026decoherence} and the Bell-measurement timing of Ref.~\cite{ali2026nearperfect}, the variable here is the placement of a heralded reversal under an explicit receiver-operation constraint. Our coherence calculation quantifies the absolute recovery probability and its relative gain over full-space reversal.
\par\endgroup

The distinction from those frameworks is therefore in the constrained design question and the resulting statements, rather than in the inverse formula. To our knowledge, prior work has not solved this recovery-placement problem under the two receiver operation classes considered here. We hold the encoding, Bell basis, monitored branch, and maximal exact sender-PR family fixed, and compare two receiver operation classes. This yields (i) a closed fidelity-constrained timing rule, including its duration-dependent regime structure and endpoint degeneracy, for one preassigned timing used on every run when the receiver has only one outcome-independent filter; (ii) a pooled-fidelity time-sharing theorem for $\Gamma T\geq(\ln 3)/2$, which optimizes over every input-independent probability distribution of those timings; and (iii) an all-$T>0$ redundancy theorem at $\Gamma>0$ for positive-time sender PR in the representative model when the receiver may reverse each complete Bell branch. None of these statements is asserted outside its stated operation class and fidelity convention. The comparison thus identifies which receiver filtering capability and which operational quality constraint make an intermediate sender reversal useful.

Full-space recovery can become unnecessarily restrictive when the encoded support excludes the least-reversible carrier directions. High-dimensional quantum teleportation and multi-level quantum communication have attracted increasing attention because they can encode more information per carrier and enrich the set of accessible entangled resources \cite{wang2015quantum,luo2019quantum,hu2020experimental,kaur2024cyclic}. Experimental progress has demonstrated teleportation beyond the qubit setting, including qutrit and other high-dimensional photonic platforms \cite{wang2015quantum,luo2019quantum,hu2020experimental}. At the same time, the richer level structure of such systems also introduces additional decay channels and inter-level couplings, which makes noise mitigation more delicate than in two-level models \cite{breuer2002theory}. In practical implementations, moreover, the initial state often occupies only a proper subspace of the available Hilbert space. For example, a qutrit may encode information only in a two-level superposition, leaving the remaining level unoccupied. In that situation, a full-space reversal is not only stronger than necessary, but can also reduce the success probability by being constrained by directions that are irrelevant to the encoded state.

The support-restricted inverse remains an important ingredient. When the input occupies only a proper subspace, attempting to invert unused carrier directions can unnecessarily lower the heralding probability. We call the exact inverse restricted to the initially occupied support Partial Reversal (PR). Its structural advantage over full-space EA-WMR can be decided by the intersection of the encoded support with the least-reversible right-singular subspace. At the protocol level, however, a sender-side PR is not automatically beneficial: its value depends on what Charlie can reverse after receiving the Bell outcome. This receiver-capability dependence is the central subject of the paper.

In this work, ``qudit'' or ``multi-level carrier'' refers to a finite-dimensional quantum carrier of dimension $d$. The PR construction is formulated for an arbitrary encoded support $S$ embedded in such a carrier through the isometry $Q_S$. The explicit protocol-level calculations and numerical figures below use qutrit carriers ($d=3$) with two-dimensional encoded supports $S\subseteq\{0,1,2\}$, which provide the minimal multi-level setting in which support-dependent recovery advantages can be displayed transparently.

The contributions are as follows. First, using the established measurement-reversal bound, we formulate the restricted map $K_0Q$ and give a sharp support/full-space equality criterion that includes degenerate and full-space-singular cases. Second, for an outcome-independent full-space receiver filter, we derive the Haar-averaged heralded metrics and solve both the fixed-schedule sender-PR timing problem and, for $\Gamma T\geq(\ln 3)/2$, its randomized time-sharing extension under a pooled minimum success-weighted conditional fidelity. Third, for full Bell-branch receiver recovery, we derive the exact branch probability and prove a model-specific redundancy result: for the representative $V/\Lambda$, $S=\{0,2\}$ configuration, every positive-time maximal exact sender PR is inferior to omitting it and reversing the composite branch at the receiver. Fourth, we quantify how partial coherence between the two $V$-type decay paths changes both the absolute PR probability and its relative full-space gain. The protected-sender and whole-interval two-qubit Ornstein--Uhlenbeck calculations are retained as supporting benchmarks, not as protocol-level optimality claims.

The remainder of this paper is organized as follows. Section~\ref{sec.2} fixes the recovery baseline and qutrit teleportation convention. Section~\ref{sec.3} applies measurement reversal to an encoded support and derives the spectral comparison with full-space EA-WMR. Section~\ref{sec.4} introduces the monitored qutrit models, distinguishes the two receiver architectures, proves the fixed-schedule, time-sharing, and redundancy results, and presents their numerical illustration. Supporting derivations and the whole-interval two-qubit example are collected in the appendices. Section~\ref{sec:conclusion} concludes.

\section{Background and Qutrit Teleportation Setting}\label{sec.2}
\subsection{Full-space EA-WMR baseline}

We use the Environment-Assisted Weak Measurement Reversal (EA-WMR) scheme of Ref.~\cite{wang2014environment} as the full-space recovery baseline against which PR is compared. In an open-system Kraus decomposition $\rho_s(t)=\sum_\ell K_\ell(t)\rho_s(0)K_\ell^\dagger(t)$, environmental monitoring projects the system into a single Kraus branch \cite{kraus1971general}; we focus throughout on the zero-photon no-jump branch represented by $K_0(t)$. EA-WMR seeks to invert $K_0(t)$ on the entire Hilbert space, and its successful weak-measurement element and maximal success probability are
\begin{equation}\label{eq:ea_baseline}
    M_{\mathrm{EA}}(t)=c_{\mathrm{EA}}(t)\,K_0^{-1}(t),
    \qquad
    p_{\mathrm{EA}}^{\max}(t)=\sigma_{\min}^{2}\!\big(K_0(t)\big),
\end{equation}
defined whenever $K_0(t)$ is invertible. EA-WMR thus pays a penalty for the least-reversible right-singular subspace of $K_0(t)$ even when that subspace is irrelevant to the encoded state, which is the structural limitation that PR is designed to lift in the next section.

\subsection{Quantum Teleportation}

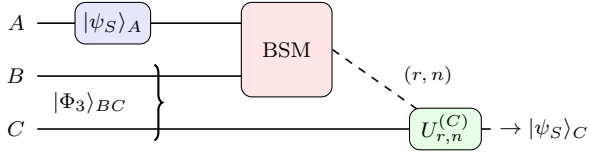
\begin{figure}[!t]
    \centering
    \begin{tikzpicture}[
        >={Latex[length=1.8mm]},
        every node/.style={font=\footnotesize},
        wlbl/.style={anchor=east, font=\footnotesize},
        op/.style={draw, rounded corners, fill=black!5, align=center,
                   minimum height=0.55cm, minimum width=0.9cm, font=\footnotesize, inner sep=2pt},
        bsm/.style={draw, rounded corners, fill=red!10, align=center,
                   minimum height=1.25cm, minimum width=1.2cm, font=\footnotesize, inner sep=2pt},
        pinput/.style={draw, rounded corners, fill=blue!10, align=center,
                   minimum height=0.55cm, minimum width=1.0cm, font=\footnotesize, inner sep=2pt},
        ccable/.style={->, dashed, semithick}
    ]
        \draw[semithick] (0,1.4) -- (3.3,1.4);
        \draw[semithick] (0,0.7) -- (3.3,0.7);
        \draw[semithick] (0,0.0) -- (6.0,0.0);
        \node[wlbl] at (-0.05,1.4) {$A$};
        \node[wlbl] at (-0.05,0.7) {$B$};
        \node[wlbl] at (-0.05,0.0) {$C$};
        \node[pinput] at (1.0,1.4) {$|\psi_S\rangle_A$};
        \draw[thick, decorate, decoration={brace, amplitude=3pt}]
            (1.55,0.85) -- (1.55,-0.15);
        \node[align=center, font=\footnotesize] at (0.7,0.35) {$|\Phi_3\rangle_{BC}$};
        \node[bsm] (BSM) at (3.3,1.05) {BSM};
        \draw[ccable] (BSM.east) -- (5.4,0.0)
            node[pos=0.55, above right, font=\scriptsize] {$(r,n)$};
        \node[op, fill=green!10, minimum width=0.95cm] (Urn) at (5.4,0.0)
            {$U_{r,n}^{(C)}$};
        \node[anchor=west, font=\footnotesize] at (6.0,0.0) {$\to|\psi_S\rangle_C$};
    \end{tikzpicture}
    \caption{Qutrit teleportation setting. Alice holds the encoded input qutrit $A$ and resource half $B$, while Charlie holds $C$; $BC$ is initialized in the maximally entangled state $|\Phi_3\rangle_{BC}$. Alice's generalized Bell measurement (BSM) on $AB$ produces the classical outcome $(r,n)$, where $r$ is the shift index and $n$ the phase index; Charlie applies the corresponding Heisenberg--Weyl correction $U_{r,n}^{(C)}$ to recover $|\psi_S\rangle_C$. The explicit Bell basis and correction convention are given in Appendix~\ref{app:qutrit_teleportation}.}
    \label{fig:teleportation_schematic}
\end{figure}

Quantum teleportation enables the remote transfer of an unknown quantum state using shared entanglement and classical channels. Originally developed for qubits \cite{bennett1993teleporting,bouwmeester1997experimental,PhysRevLett.80.1121}, high-dimensional and multi-level teleportation protocols can encode more information per carrier and have been experimentally demonstrated in photonic systems \cite{luo2019quantum,hu2020experimental}. The standard high-dimensional teleportation flow---input qudit on $A$, entangled qudit pair on $BC$, Bell-basis measurement on $AB$, classical communication to Charlie, and a unitary correction $U_{r,n}$ on $C$---is summarized in Fig.~\ref{fig:teleportation_schematic}; Appendix~\ref{app:qutrit_teleportation} lists the explicit qutrit Bell basis and Heisenberg--Weyl correction operators used in this paper. We reserve $m$ for the encoded-support dimension and use $(r,n)$ for the Bell shift and phase indices, respectively.

Multi-level systems, however, tend to be more sensitive to decoherence because their richer level structure allows additional decay channels and unwanted transitions induced by system--environment interactions \cite{wang2015quantum}. Such processes can cause energy loss or leakage to undesired levels, thereby degrading teleportation fidelity. Under ideal conditions, the standard teleportation process preserves the quantum state perfectly. In realistic noisy environments, however, non-unitary evolution can drive the teleported state away from its intended form. This motivates the need for flexible and efficient state-recovery mechanisms to counteract these detrimental effects and enhance transmission reliability.

For a multi-level carrier, recovery can be tailored to its initially occupied support. In the following sections, we use Partial Reversal (PR) to compare recovery placement under two receiver operation classes.

For concreteness, the multi-level teleportation protocol analyzed in this work involves three qutrits, denoted $A$, $B$, and $C$, with single-system Hilbert spaces
\begin{equation}\label{eq:ABC_hilbert_spaces}
    \mathcal{H}_A=\mathcal{H}_B=\mathcal{H}_C=\mathbb{C}^3.
\end{equation}
Alice holds the input qutrit $A$, prepared in an unknown encoded state on a two-level support $S\subseteq\{0,1,2\}$:
\begin{equation}\label{eq:input_state_A}
    |\psi_S\rangle_A=\sum_{j\in S}a_j|j\rangle_A,
    \qquad
    \proj_A^{S}=\sum_{j\in S}|j\rangle\langle j|.
\end{equation}
The shared entangled resource is the maximally entangled qutrit pair on $BC$,
\begin{equation}\label{eq:phi3_resource}
    |\Phi_3\rangle_{BC}=\frac{1}{\sqrt{3}}\sum_{k=0}^{2}|k\rangle_B|k\rangle_C,
\end{equation}
with Alice holding the resource half $B$ and Charlie holding $C$, so that Alice can perform the joint Bell measurement on $AB$. Throughout this paper, the labels $\Lambda$-type, $V$-type, and all-transition refer to local jump-operator topologies that act on a single qutrit (the sender's $A$ or the receiver's $C$). They are noise topologies and do not designate distinct teleportation participants. Each topology is modeled by one coherent Lindblad jump operator whose displayed pathways are summed before forming $L^{\mu\dagger}L^\mu$; it is therefore not equivalent to a collection of independently distinguishable jump channels. ``All-transition'' here means the three directed transitions in the chosen orientation of the level graph, not all six bidirectional transitions between the three basis states. Their explicit matrix forms in the computational basis $\{|0\rangle,|1\rangle,|2\rangle\}$ are given in Appendix~\ref{app:jump_operators} and compared in Section~\ref{subsec:model_compare}.

\begin{figure*}[!t]
    \centering
    \begin{tikzpicture}[
        >={Latex[length=2mm]},
        every node/.style={font=\footnotesize},
        block/.style={draw, rounded corners, align=center,
                      minimum height=0.85cm, minimum width=2.05cm,
                      font=\footnotesize, inner sep=2pt},
        blueblock/.style={draw, rounded corners, fill=blue!10, align=center,
                      minimum height=0.85cm, minimum width=2.05cm,
                      font=\footnotesize, inner sep=2pt},
        redblock/.style={draw, rounded corners, fill=red!10, align=center,
                      minimum height=0.85cm, minimum width=2.05cm,
                      font=\footnotesize, inner sep=2pt},
        greenblock/.style={draw, rounded corners, fill=green!10, align=center,
                      minimum height=0.85cm, minimum width=2.05cm,
                      font=\footnotesize, inner sep=2pt},
        line/.style={->, semithick}
    ]
        \node[blueblock] (input) at (0,0)   {encoded input\\ on $\mathcal{H}_{\mathrm{enc}}$};
        \node[block]     (k0)    at (2.8,0) {no-jump\\ $K_0(t)$};
        \node[block]     (b)     at (5.6,0) {restricted\\ $B(t)=K_0(t)Q$};
        \node[redblock]  (rpr)   at (8.5,0) {PR inverse\\ $R_{\mathrm{PR}}=Q B^{+}(t)$};
        \node[redblock]  (mpr)   at (11.4,0){success branch\\ $M_{\mathrm{PR}}=c\,R_{\mathrm{PR}}$};
        \node[greenblock](out)   at (14.3,0){restored\\ on $\Pi$};
        \draw[line] (input) -- (k0);
        \draw[line] (k0)    -- (b);
        \draw[line] (b)     -- (rpr);
        \draw[line] (rpr)   -- (mpr);
        \draw[line] (mpr)   -- (out);
        \node[font=\scriptsize, anchor=center] at (12.85,0.55)
            {$R_{\mathrm{PR}}K_0(t)\Pi=\Pi$};
    \end{tikzpicture}
    \caption{Support-restricted PR module. The no-jump Kraus branch $K_0(t)$ is restricted to the encoded support through $B(t)=K_0(t)Q$, where $Q$ is an isometry into $\mathcal{H}_{\mathrm{enc}}$. When $B(t)$ has full column rank, the PR inverse $R_{\mathrm{PR}}=Q B^{+}(t)$ satisfies $R_{\mathrm{PR}}K_0(t)\Pi=\Pi$, and the physical success branch is $M_{\mathrm{PR}}=c R_{\mathrm{PR}}$ with $c$ fixed by the POVM constraint $M_{\mathrm{PR}}^\dagger M_{\mathrm{PR}}\le I$. Successful postselection of this module restores the encoded state. In the unprotected-sender variant of Architecture A, an additional residual sender evolution $K_A^\mu(s)$ acts after this PR step, so the final teleportation output is generally not exactly the input.}
    \label{fig:pr_scheme}
\end{figure*}
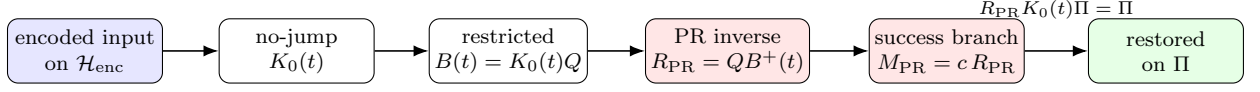

\begin{samepage}
\section{Partial Reversal (PR)}\label{sec.3}

We apply the established probabilistic measurement-reversal construction to the domain that actually carries the encoded information. The support-restricted recovery module shown in Fig.~\ref{fig:pr_scheme} is a Moore--Penrose left inverse of the restricted no-jump map. In many multi-level protocols, the initial state occupies only a proper subspace of the carrier Hilbert space. A full-state reversal then imposes unnecessary constraints from unoccupied directions. We use ``Partial Reversal'' (PR) for the exact probabilistic reversal defined only on this encoded domain.
\par\end{samepage}

To describe the postselected dynamics, we use the propagation-method picture \cite{ding2021stochastic}, in which a measurement on the environment selects a Kraus branch associated with a given photon record. In particular, the zero-photon branch is represented by the no-jump Kraus operator $K_0(t)$. If the initial state is $\rho_s(0)$, then conditioning on zero-photon detection gives the state
\begin{equation}\label{eq:pr_conditional_state}
    \rho_1(t)=\frac{K_0(t)\rho_s(0)K_0^\dagger(t)}{p_1(t)},
\end{equation}
with conditional probability
\begin{equation}\label{eq:pr_p1}
    p_1(t)=\operatorname{Tr}\!\left[K_0(t)\rho_s(0)K_0^\dagger(t)\right].
\end{equation}

We now introduce the encoded subspace $\mathcal{H}_{\mathrm{enc}} \subset \mathcal{H}$ and its projector $\Pi$, defined by
\begin{equation}\label{eq:pr_support_condition}
    \rho_s(0)=\Pi \rho_s(0)\Pi .
\end{equation}
All physically relevant initial states are supported on $\mathcal{H}_{\mathrm{enc}}$. The PR construction should therefore restore the action of $K_0(t)$ only on this subspace.

Let $m=\rank(\Pi)$ and let $Q:\mathbb{C}^{m}\to\mathcal{H}$ be an isometry whose columns form an orthonormal basis of $\mathcal{H}_{\mathrm{enc}}$, so that
\begin{equation}\label{eq:pr_Q_relations}
    Q^\dagger Q=I_m,
    \qquad
    QQ^\dagger=\Pi.
\end{equation}
The restricted no-jump map is represented by the rectangular matrix
\begin{equation}\label{eq:pr_B_def}
    B(t)=K_0(t)Q\in\mathcal{L}(\mathbb{C}^m,\mathcal{H}).
\end{equation}
Exact PR on the encoded support exists if and only if $\rank B(t)=m$. When this holds, the canonical PR operator is the Moore--Penrose pseudoinverse of the restricted map composed with $Q$,
\begin{equation}\label{eq:pr_left_inverse}
    R_{\mathrm{PR}}(t)
    =
    Q\,B^{+}(t)
    =
    Q\bigl[B^\dagger(t)B(t)\bigr]^{-1}B^\dagger(t).
\end{equation}
A direct calculation using Eq.~\eqref{eq:pr_Q_relations} verifies the defining left-inverse property
\begin{align}\label{eq:pr_identity_on_support}
    R_{\mathrm{PR}}(t)\,K_0(t)\,\Pi
    &=
    Q\bigl[B^\dagger(t)B(t)\bigr]^{-1}B^\dagger(t)B(t)Q^\dagger
    \nonumber\\
    &=QQ^\dagger=\Pi.
\end{align}
Equivalently, $R_{\mathrm{PR}}(t)$ is the Moore--Penrose pseudoinverse of the restricted map $K_0(t)\Pi:\mathcal{H}_{\mathrm{enc}}\to\mathcal{H}$, with the inverse understood only on the encoded support.

The general singular-value limit for exact probabilistic reversal is known~\cite{nielsen1997reversible,nielsen1998information,cheong2012balance}. Here it is applied to the rectangular, support-restricted branch $B(t)=K_0(t)Q$, rather than to the full carrier operator or to a hand-picked principal submatrix. The classification of all left inverses, including non-canonical ones, is summarized in Appendix~\ref{app:pr_pseudoinverse_construction}; the operator in Eq.~\eqref{eq:pr_left_inverse} is the Moore--Penrose minimum-norm choice and is the only one used below.

The minimal operator norm and the resulting maximal PR success probability are
\begin{equation}\label{eq:pr_norm_bound}
    \|R_{\mathrm{PR}}(t)\|_{\mathrm{op}}
    =
    \frac{1}{\sigma_{\min}(B(t))},
    \qquad
    p_{\mathrm{PR}}^{\max}(t)=\sigma_{\min}^2(B(t)),
\end{equation}
where $\sigma_{\min}(B(t))$ is the smallest singular value of the restricted matrix in Eq.~\eqref{eq:pr_B_def}. Among physical branches that reverse this restricted map exactly, the optimal scaling is consequently $c_*(t)=\sigma_{\min}(B(t))$.

\begin{corollary}[Exact restoration on PR success]
\label{cor:exact_restoration}
Suppose the initial state satisfies Eq.~\eqref{eq:pr_support_condition} and the restricted inverse in Eq.~\eqref{eq:pr_left_inverse} exists. Let the successful weak-measurement element be
\begin{equation}\label{eq:pr_success_measurement}
    M_{\mathrm{PR}}(t)=c(t)\,R_{\mathrm{PR}}(t),
\end{equation}
where $c(t)>0$ is chosen such that
\begin{equation}\label{eq:pr_povm_condition}
    M_{\mathrm{PR}}^\dagger(t)M_{\mathrm{PR}}(t)\le I.
\end{equation}
Then successful PR restores the initial state exactly, and the total success probability is
\begin{equation}\label{eq:pr_total_success_general}
    p_{\mathrm{tot}}(t)=c^2(t).
\end{equation}
\end{corollary}

\paragraph*{Proof.}
After the zero-photon measurement, the state is given by Eq.~\eqref{eq:pr_conditional_state}. Applying the successful measurement branch $M_{\mathrm{PR}}(t)$ yields
\begin{equation}\label{eq:pr_state_after_success}
    \rho_2(t)=\frac{M_{\mathrm{PR}}(t)\rho_1(t)M_{\mathrm{PR}}^\dagger(t)}{p_2(t)},
\end{equation}
where
\begin{equation}\label{eq:pr_p2}
    p_2(t)=\operatorname{Tr}\!\left[M_{\mathrm{PR}}(t)\rho_1(t)M_{\mathrm{PR}}^\dagger(t)\right].
\end{equation}
Using Eqs.~\eqref{eq:pr_support_condition}, \eqref{eq:pr_identity_on_support}, and \eqref{eq:pr_success_measurement}, we obtain
\begin{equation}\label{eq:pr_recovery_numerator}
    M_{\mathrm{PR}}(t)K_0(t)\rho_s(0)K_0^\dagger(t)M_{\mathrm{PR}}^\dagger(t)
    =
    c^2(t)\rho_s(0).
\end{equation}
Hence,
\begin{equation}\label{eq:pr_recovery_exact}
    \rho_2(t)=\rho_s(0),
\end{equation}
and the second-step conditional success probability becomes
\begin{equation}\label{eq:pr_p2_explicit}
    p_2(t)=\frac{c^2(t)}{p_1(t)}.
\end{equation}
Therefore the total success probability is
\begin{equation}\label{eq:pr_total_success}
    p_{\mathrm{tot}}(t)=p_1(t)p_2(t)=c^2(t).
\end{equation}
\hfill $\square$

To complete the weak-measurement description, one may introduce a failure operator
\begin{equation}\label{eq:pr_failure_operator}
    M_f(t)=\sqrt{I-M_{\mathrm{PR}}^\dagger(t)M_{\mathrm{PR}}(t)},
\end{equation}
so that the successful and failed branches together form a valid two-outcome POVM. The maximal scaling that satisfies Eq.~\eqref{eq:pr_povm_condition} is $c_*(t)=\|R_{\mathrm{PR}}(t)\|_{\mathrm{op}}^{-1}=\sigma_{\min}(B(t))$, which reproduces the maximal PR success probability $p_{\mathrm{PR}}^{\max}(t)=\sigma_{\min}^2(B(t))$ already stated in Eq.~\eqref{eq:pr_norm_bound}.

Comparing this with EA-WMR makes the structural distinction transparent. In the standard full-space approach, one attempts to invert $K_0(t)$ on the entire Hilbert space, provided $K_0(t)$ is invertible, with maximal success probability
\begin{equation}\label{eq:ea_probability_full}
    p_{\mathrm{EA}}^{\max}(t)=\sigma_{\min}^2\!\big(K_0(t)\big).
\end{equation}
Because the PR construction minimizes only over the encoded subspace rather than the full Hilbert space, whenever the full-space EA-WMR inverse exists,
\begin{equation}\label{eq:pr_geq_ea}
    p_{\mathrm{PR}}^{\max}(t)\ge p_{\mathrm{EA}}^{\max}(t).
\end{equation}
Let $E_{\min}(t)$ denote the right-singular subspace of $K_0(t)$ associated with $\sigma_{\min}(K_0(t))$. Equality holds precisely when $\mathcal H_{\mathrm{enc}}\cap E_{\min}(t)\neq\{0\}$; if the intersection is trivial, the inequality is strict. This eigenspace formulation also covers a degenerate smallest singular value without selecting an arbitrary ``least-reversible direction.'' If $K_0(t)$ is singular on the full Hilbert space but $B(t)=K_0(t)Q$ remains full rank on the encoded support, full-space EA-WMR is not defined whereas PR is well defined and yields a finite $p_{\mathrm{PR}}^{\max}(t)$. The PR construction therefore strictly extends the operating regime of full-space EA-WMR.

\paragraph*{Remark (spectral criterion for PR gain).}
For a fixed encoded projector $\Pi$ and invertible $K_0(t)$, an exact measure of the PR advantage is
\begin{equation}\label{eq:gain_ratio_exact}
    \mathcal{G}_{\Pi}(t)
    =
    \frac{p_{\mathrm{PR}}^{\max}(t)}{p_{\mathrm{EA}}^{\max}(t)}
    =
    \frac{\sigma_{\min}^2\!\big(B(t)\big)}{\sigma_{\min}^2\!\big(K_0(t)\big)}\ge 1.
\end{equation}
In the notation above, equality is equivalent to $\mathcal H_{\mathrm{enc}}\cap E_{\min}(t)\neq\{0\}$. PR is strictly advantageous when this intersection is trivial, because the restriction then has a strictly larger minimum Rayleigh quotient, and hence a larger restricted smallest singular value. Equation~\eqref{eq:gain_ratio_exact} provides a sharp, model-independent diagnostic for when support-restricted reversal yields a substantial improvement over the full-space inverse. If $K_0(t)$ is singular, the ratio itself is not used; instead one compares the existence and success factor of PR directly with the failure of an exact full-space inverse.

\paragraph*{Numerical construction.}
We implement PR as follows. First, we compute an isometry $Q$ whose columns form an orthonormal basis of $\mathcal{H}_{\mathrm{enc}}$. We then form the restricted matrix $B(t)=K_0(t)Q$ and compute its thin singular-value decomposition $B(t)=U\Sigma V^\dagger$. If the smallest singular value satisfies $\sigma_{\min}(B(t))\le \varepsilon_{\mathrm{svd}}$, where $\varepsilon_{\mathrm{svd}}$ is the prescribed numerical tolerance, then exact PR on the chosen support is either impossible or numerically unstable. Otherwise, we set
\begin{equation}\label{eq:pr_numerical_implementation}
    R_{\mathrm{PR}}(t)=QV\Sigma^{-1}U^\dagger,
    \qquad
    p_{\mathrm{PR}}^{\max}(t)=\sigma_{\min}^2(B(t)).
\end{equation}
In practice, one should not attempt to invert the full matrix $\Pi K_0^\dagger(t)K_0(t)\Pi$ directly, because it is singular on $(I-\Pi)\mathcal{H}$ whenever $\Pi\neq I$. The stable implementation works with the compressed Gram matrix $Q^\dagger K_0^\dagger(t)K_0(t)Q=B^\dagger(t)B(t)$ or, equivalently, with the thin SVD of $B(t)$. A reliable default is $\varepsilon_{\mathrm{svd}}=\max(d,m)\,\epsilon_{\mathrm{mach}}\,\sigma_{\max}(B(t))$, where $d=\dim\mathcal{H}$ and $m=\rank(\Pi)$. This notation is kept distinct from the physical sender-intervention time $\tau$ introduced in Section~\ref{subsec:fixed_timing}.

\subsection{Projector construction for qutrit initial supports}\label{subsec:qutrit_supports}

The projector language avoids basis ambiguity and provides a compact description of PR in a qutrit system. For a qutrit with computational basis $\{|0\rangle,|1\rangle,|2\rangle\}$, the three relevant two-level supports are
\begin{equation}\label{eq:qutrit_projectors}
\begin{aligned}
    \Pi_{01}&=|0\rangle\langle 0|+|1\rangle\langle 1|, \\
    \Pi_{02}&=|0\rangle\langle 0|+|2\rangle\langle 2|, \\
    \Pi_{12}&=|1\rangle\langle 1|+|2\rangle\langle 2|.
\end{aligned}
\end{equation}
For example, if the initial state is
\begin{equation}\label{eq:qutrit_state_01}
    |\psi_{01}\rangle=\alpha |0\rangle+\beta |1\rangle,\qquad |\alpha|^2+|\beta|^2=1,
\end{equation}
then its density matrix satisfies
\begin{equation}
    \rho_{01}(0)=|\psi_{01}\rangle\langle\psi_{01}|=\Pi_{01}\rho_{01}(0)\Pi_{01}.
\end{equation}
The PR operator is therefore constructed from $\Pi_{01}$ through Eq.~\eqref{eq:pr_left_inverse}. The same procedure applies to states supported on $\Pi_{02}$ and $\Pi_{12}$.

In this way, PR is not tied to a specific matrix ordering or ad hoc truncation rule. Instead, it is determined entirely by the support projector $\Pi$ associated with the encoded subspace. The corresponding support dependence of $\sigma_{\min}^2(B(t))$ for the three qutrit topologies $\mu\in\{\Lambda,V,\mathrm{all}\}$ defined in Appendix~\ref{app:jump_operators} is the direct numerical content of Fig.~\ref{fig:sender_support_advantage} of Section~\ref{subsec:numerical_support_advantage}.

For later use in the qutrit protocols of Section~\ref{sec.4}, we introduce the shorthand
\begin{equation}\label{eq:sigma_min_S_def}
    \sigma_{\min}^{S}(K)
    :=
    \sigma_{\min}\!\bigl(K\,Q_S\bigr)
    =
    \lambda_{\min}\!\bigl(Q_S^\dagger K^\dagger K\,Q_S\bigr)^{1/2},
\end{equation}
where $Q_S:\mathbb{C}^{|S|}\to\mathcal{H}$ is any isometry whose range is the encoded support $\mathcal{H}_S$, so that $\Pi_S=Q_S Q_S^\dagger$. The notation $\sigma_{\min}^{S}(K)$ always denotes the smallest singular value of the restricted map $K:\mathcal{H}_S\to\mathcal{H}$, equivalently the smallest singular value of the rectangular matrix $K Q_S\in\mathcal{L}(\mathbb{C}^{|S|},\mathcal{H})$. It is \emph{not} the smallest singular value of the full operator $K\Pi_S$ acting on all of $\mathcal{H}$, which would always be zero whenever $\Pi_S\neq I$.

\section{PR-Assisted Noisy Qutrit Teleportation Protocols}\label{sec.4}

\subsection{Model-dependent comparison between PR and EA-WMR}\label{subsec:model_compare}

To illustrate the structural origin of the PR advantage, we compare PR and EA-WMR under several representative dissipative couplings. For a Markovian no-jump branch generated by a system Hamiltonian $H_{\mathrm{sys}}$ and a jump operator $L$, the corresponding no-jump Kraus operator is
\begin{equation}\label{eq:k0_markovian_general}
    K_0(t)=\exp\!\left[\left(-iH_{\mathrm{sys}}-\frac{\Gamma}{2}L^\dagger L\right)t\right].
\end{equation}
The difference between PR and EA-WMR is then controlled by whether the encoded subspace intersects the least-reversible right-singular subspace of $K_0(t)$.

A complementary short-time interpretation can be given in terms of the effective no-jump decay operator
\begin{equation}\label{eq:decay_operator_def}
    D=\Gamma L^\dagger L,
\end{equation}
or, more generally, $D=\sum_\mu \Gamma_\mu L_\mu^\dagger L_\mu$ for multiple dissipative channels. Since
\begin{equation}\label{eq:short_time_KdK}
    K_0^\dagger(t)K_0(t)=I-Dt+O(t^2),
\end{equation}
one finds
\begin{subequations}\label{eq:short_time_decay_criterion}
\begin{align}
    p_{\mathrm{EA}}^{\max}(t)
    &=
    1-t\,\lambda_{\max}(D)+O(t^2),
    \label{eq:short_time_EA}
    \\
    p_{\mathrm{PR}}^{\max}(t)
    &=
    1-t\,\lambda_{\max}\!\left(\Pi D\Pi\big|_{\mathcal{H}_{\mathrm{enc}}}\right)+O(t^2).
    \label{eq:short_time_PR}
\end{align}
\end{subequations}
The short-time expansions in Eqs.~\eqref{eq:short_time_EA} and~\eqref{eq:short_time_PR} show that PR has a strict first-order success-probability advantage exactly when $\lambda_{\max}\!\left(\Pi D\Pi\big|_{\mathcal{H}_{\mathrm{enc}}}\right)<\lambda_{\max}(D)$, i.e., when the largest decay rate on the encoded support is smaller than its full-space counterpart. In non-diagonal models, the highest-decay directions of $D$ correspond to bright decay eigenmodes rather than to individual bare energy levels.

For the purpose of comparison, we consider three local single-qutrit jump-operator topologies, $\mu\in\{\Lambda,V,\mathrm{all}\}$, defined explicitly in Appendix~\ref{app:jump_operators}. Each topology is represented by a single coherent jump operator, rather than by separate independently monitored channels for its constituent pathways. They are local noise topologies acting on the same qutrit Hilbert space $\mathbb{C}^3$, not different teleportation participants. The two-qubit collective example used as the supporting analytic illustration in Section~\ref{subsec:two_qubit_supporting} is a separate reference model.

A subtle point is worth emphasizing. Diagonal $L^\dagger L$ does not by itself guarantee equality between PR and EA-WMR for every encoded support. PR coincides with EA-WMR only when the encoded support intersects the full-space least-reversible right-singular subspace. For the $\Lambda$-type convention used here, $L_\Lambda^\dagger L_\Lambda=2|0\rangle\langle 0|$ and that subspace is the one-dimensional span of $|0\rangle$, so any support containing $|0\rangle$ saturates the equality in Eq.~\eqref{eq:pr_geq_ea}. A support such as $S=\{1,2\}$ that excludes this direction can still yield a strict PR advantage even for the $\Lambda$-type model. For the $V$-type and all-transition topologies, $L^\dagger L$ is non-diagonal in the computational basis and its bright decay eigenmodes are not aligned with any single basis vector, so the support dependence of the PR advantage is governed by the geometric overlap between the encoded support and those bright eigenmodes rather than by bare energy-level separations. The numerical illustration of this support-dependent PR advantage is shown in Fig.~\ref{fig:sender_support_advantage} of Section~\ref{subsec:numerical_support_advantage}.

\subsection{Dependence on coherence between decay paths}\label{subsec:path_coherence}

The single $V$-type jump operator used above represents fully coherent, indistinguishable decay pathways, as can arise from coupling two transitions to a common structured reservoir~\cite{yannopapas2009plasmon}. To separate this assumption from the use of a qutrit itself, let
\begin{equation}
    J_0=|2\rangle\langle0|,
    \qquad
    J_1=|2\rangle\langle1|,
\end{equation}
and introduce a real cross-damping parameter $0\leq\eta\leq1$. The positive Kossakowski matrix and the corresponding dimensionless decay matrix are
\begin{align}
    C_\eta&=
    \begin{pmatrix}1&\eta\\ \eta&1\end{pmatrix},
    \label{eq:coherence_matrix}
    \\
    \widetilde D_\eta&=
    |0\rangle\langle0|+|1\rangle\langle1|
    +\eta\bigl(|0\rangle\langle1|+|1\rangle\langle0|\bigr),
    \label{eq:coherence_decay_operator}
    \\
    K_\eta(t)&=\exp[-\Gamma t\widetilde D_\eta/2].
    \label{eq:coherence_no_jump}
\end{align}
Thus the rate-weighted short-time operator of Eq.~\eqref{eq:short_time_KdK} is $D=\Gamma\widetilde D_\eta$. Equivalently, the two diagonalized jump operators are
$J_\pm=\sqrt{(1\pm\eta)/2}\,(J_0\pm J_1)$. Thus $\eta=1$ reproduces the single coherent operator $L^V=J_0+J_1$, whereas $\eta=0$ gives the same no-jump dynamics as two independent equal-rate pathways.

For $S=\{0,2\}$, direct diagonalization of $\widetilde D_\eta$ gives the maximal sender-side recovery factors
\begin{align}
    p_{\mathrm{PR}}^{02}(\eta,t)
    &=e^{-\Gamma t}\cosh(\eta\Gamma t),
    \label{eq:coherence_pPR}
    \\
    p_{\mathrm{EA}}(\eta,t)
    &=e^{-(1+\eta)\Gamma t},
    \label{eq:coherence_pEA}
    \\
    \mathcal G_{02}(\eta,t)
    &=\frac{p_{\mathrm{PR}}^{02}(\eta,t)}{p_{\mathrm{EA}}(\eta,t)}
    =\frac{1+e^{2\eta\Gamma t}}{2}.
    \label{eq:coherence_gain}
\end{align}
The absolute sender-side factor and the relative gain convey different physics. At $\eta=0$ the support advantage disappears, $\mathcal G_{02}=1$. At $\eta=1$, an exact dark mode produces $p_{\mathrm{PR}}^{02}\to1/2$ as $t\to\infty$. For every fixed $\eta<1$, however, $p_{\mathrm{PR}}^{02}\to0$ even though the ratio in Eq.~\eqref{eq:coherence_gain} can grow. A large relative gain therefore does not imply robust long-time sender-side recovery success under imperfect pathway coherence.

\subsection{Protected-sender Architecture-A benchmark}\label{subsec:architecture_A_protected}

We now place PR inside a noisy qutrit teleportation protocol. We adopt a \emph{local-noise model} in which Alice's qutrit $A$ and Charlie's receiver qutrit $C$ each interact with their own Markovian bath, with no-jump Kraus operators $K_A^{\mu}(t)$ and $K_C^{\chi}(t)$, and the entangled partner $B$ is taken to be ideally protected. The labels $\mu,\chi\in\{\Lambda,V,\mathrm{all}\}$ index the local single-qutrit jump-operator topologies of Section~\ref{subsec:model_compare}. Alice's encoded input lies on a two-level support $S\subseteq\{0,1,2\}$, $|S|=m=2$, with projector $\Pi_A^S=\sum_{j\in S}|j\rangle\langle j|$. A block-level extension in which the sender $AB$ shares a collective bath with no-jump propagator $K_{AB,0}(t)$ and joint encoded support $\Pi_{AB}$ follows from the same singular-value argument and is recorded for completeness in Appendix~\ref{app:block_extension}.

In compact symbolic form, the two timing variants compared in this section are
\begin{align*}
\text{Protected sender:}\ &
A:\ K_A^{\mu}(t)\to\mathrm{PR}\to\text{stored},\\
&C:\ K_C^{\chi}(T),
\\
\text{Unprotected sender:}\ &
A:\ K_A^{\mu}(t)\to\mathrm{PR}\to K_A^{\mu}(s),\\
&C:\ K_C^{\chi}(T),
\end{align*}
with $T=t+s$ on the receiver. The standalone PR module that appears as the $\mathrm{PR}$ step in both timing variants is the schematic of Fig.~\ref{fig:pr_scheme}.

In the protected-sender Architecture-A benchmark, support-restricted PR is applied to $A$ at time $t$ and the sender is then \emph{protected}. Either Alice immediately Bell-measures $AB$ and stores the classical result until the receiver operation at $T=t+s$, or she decouples and stores $A$ noiselessly until a Bell measurement at $T$. In both implementations, no further dissipation acts on $A$ after PR. The receiver $C$ continues to evolve under $K_C^{\chi}$ over the full interval $[0,T]$. Operationally, the receiver branch is interpreted as follows: $C$ first undergoes the no-jump evolution $K_C^{\chi}(T)$; the environment record selects the no-jump branch; an outcome-independent full-space receiver EA-WMR element $M_{C,\mathrm{EA}}^{\chi}(T)$ is then applied; finally, after Alice's Bell outcome is available, Charlie applies the corresponding Heisenberg--Weyl correction. With this stipulated receiver architecture, the receiver factor satisfies $M_{C,\mathrm{EA}}^{\chi}(T)K_C^{\chi}(T)=c_C^{\chi}(T)\,I_C$, and combining the support-restricted PR identity on $A$ with the full-space EA-WMR identity on $C$, the total success probability factorizes and is maximized, within this architecture, by
\begin{equation}\label{eq:async_total_probability_optimal}
    p_{\mathrm{tot,max}}^{(\mathrm{prot})}(t,s)
    =
    \bigl[\sigma_{\min}^{S}\!\big(K_A^{\mu}(t)\big)\bigr]^{2}\,
    \sigma_{\min}^2\!\big(K_C^{\chi}(t+s)\big).
\end{equation}
The sender factor is the restricted singular-value factor $[\sigma_{\min}^{S}(K_A^{\mu}(t))]^2$, computed on the encoded support of Alice's input, while the receiver factor is the full-space EA-WMR factor over the total receiver interval $T=t+s$. Because exact recovery is achieved on the encoded support after PR in this benchmark, the conditional fidelity satisfies $F_{\mathrm{cond}}=1$ on success and the success-weighted effective fidelity coincides with the success probability,
\begin{equation}\label{eq:protocol_I_eff}
    \bar F_{\mathrm{eff}}^{(\mathrm{prot})}(t,s)=p_{\mathrm{tot,max}}^{(\mathrm{prot})}(t,s).
\end{equation}
We call this ordering \emph{Architecture A}: Charlie uses one outcome-independent full-space filter and then the standard Bell-dependent unitary correction. The qualifier ``within this architecture'' is essential. If the receiver filter is delayed until the Bell outcome $(r,n)$ is known and is allowed to adapt to its shifted encoded support, the protected-sender receiver factor becomes
\begin{equation}\label{eq:receiver_outcome_adaptive}
    p_{C,\mathrm{adapt}}(T)
    =\frac{1}{3}\sum_{r=0}^{2}
    \sigma_{\min}^{2}\!\bigl(K_C^\chi(T)X_C^rQ_S\bigr)
    \geq \sigma_{\min}^{2}\!\bigl(K_C^\chi(T)\bigr),
\end{equation}
where the phase index $n$ drops out because $Z_C^{-n}$ acts unitarily within a computational-basis support. This is the zero-residual-noise limit of \emph{Architecture B}, the full Bell-branch recovery analyzed in Section~\ref{subsec:full_branch_receiver}; for an unprotected sender the residual sender map must also be included inside each effective branch and Eq.~\eqref{eq:receiver_outcome_adaptive} no longer factorizes from it.
A heatmap of Eq.~\eqref{eq:protocol_I_eff} on the $(\Gamma t,\Gamma s)$ plane for a representative $V$-type sender / $\Lambda$-type receiver configuration with encoded support $\Pi_{02}$ is shown in Fig.~\ref{fig:supp_sender_diagnostics}(a) of Section~\ref{subsec:numerical_protected}.

\subsection{Architecture A: unprotected sender and a fixed receiver filter}\label{subsec:architecture_A_unprotected}

The protected-sender benchmark of Section~\ref{subsec:architecture_A_protected} provides a clean upper bound on the heralded performance under Architecture A, but it sets aside what happens if the sender continues to interact with its environment before the Bell measurement. We therefore let the sender qutrit $A$ be exposed to noise both before and after PR. The sender $A$ and receiver $C$ carry local no-jump branches $K_A^{\mu}(\cdot)$ and $K_C^{\chi}(\cdot)$, while the entangled-pair partner $B$ is ideal, with branch $I_B$. With $T=t+s$, if no operation were inserted at time $t$ the Markovian sender no-jump factors would compose across the two intervals, giving the reference propagator
\begin{equation}\label{eq:local_noise_model}
    \mathcal{K}_{\mathrm{local}}(t,s)
    =
    K_A^{\mu}(s)K_A^{\mu}(t)\otimes I_B\otimes K_C^{\chi}(T).
\end{equation}
In the unprotected-sender variant, however, a PR operation is inserted on $A$ at time $t$, so the actual successful branch is not $\mathcal{K}_{\mathrm{local}}(t,s)$ but the operator $L_{\mathrm{unprot}}$ given in Eq.~\eqref{eq:Lunprot_def} below. The encoded input on Alice's qutrit is supported on a two-level subspace,
\begin{equation}
    |\psi_S\rangle_A=\sum_{j\in S}a_j|j\rangle_A,
    \qquad
    \proj_A^S=\sum_{j\in S}|j\rangle\langle j|,
\end{equation}
with $S\in\{\{0,1\},\{0,2\},\{1,2\}\}$ and $m=|S|=2$.

In the unprotected scenario, the support-restricted PR operator $M_{A,\mathrm{PR}}^{\mu,S}(t)$ is applied at time $t$ to the sender, after which $A$ continues to evolve under the monitored no-jump branch $K_A^{\mu}(s)$ before the ideal Bell measurement. Thus success is also conditioned on a no-jump record during the residual interval $[t,T]$, whose state-dependent norm is retained explicitly below through $G_S^\mu(s)$. Meanwhile, the receiver $C$ undergoes its own evolution up to $T$ and is corrected by the EA-WMR operator $M_{C,\mathrm{EA}}^{\chi}(T)$ before the Heisenberg--Weyl correction is applied once Alice's Bell outcome is known, in the same operational order as in the protected-sender Architecture-A benchmark. The successful conditional branch is therefore generated by
\begin{align}\label{eq:Lunprot_def}
    L_{\mathrm{unprot}}(t,s)
    =\;&
    \big[K_A^{\mu}(s)M_{A,\mathrm{PR}}^{\mu,S}(t)K_A^{\mu}(t)\big]
    \nonumber\\
    &\otimes I_B\otimes
    \big[M_{C,\mathrm{EA}}^{\chi}(T)K_C^{\chi}(T)\big].
\end{align}
Using the support-restricted PR identity
\begin{equation}\label{eq:Aside_PR_identity}
    M_{A,\mathrm{PR}}^{\mu,S}(t)K_A^{\mu}(t)\proj_A^S=c_A^{\mu,S}(t)\proj_A^S,
\end{equation}
together with the full-space EA-WMR identity
\begin{equation}\label{eq:Cside_EA_identity}
    M_{C,\mathrm{EA}}^{\chi}(T)K_C^{\chi}(T)=c_C^{\chi}(T)\,I_C,
\end{equation}
the action of $L_{\mathrm{unprot}}$ on any encoded input becomes
\begin{align}\label{eq:Lunprot_on_support}
    L_{\mathrm{unprot}}(t,s)&\big(\proj_A^S\otimes I_B\otimes I_C\big)
    \nonumber\\
    =\;&
    c_A^{\mu,S}(t)c_C^{\chi}(T)\big[K_A^{\mu}(s)\proj_A^S\big]\otimes I_B\otimes I_C.
\end{align}
Equation~\eqref{eq:Lunprot_on_support} is the central identity for the unprotected sender under Architecture A: the early PR step cancels the noisy evolution that occurred \emph{before} time $t$, but the residual sender-side evolution $K_A^{\mu}(s)$ is \emph{not} reversed by the fixed receiver filter. Exact recovery within this architecture survives only when the residual support map is proportional to a known correctable isometry. In general, it produces the following heralded success--fidelity tradeoff.

To this end, define the support-restricted residual operators
\begin{align}
    G_S^{\mu}(s)
    &=
    \proj_A^S\,K_A^{\mu\dagger}(s)K_A^{\mu}(s)\,\proj_A^S,
    \label{eq:GS_def}
    \\
    M_S^{\mu}(s)
    &=
    \proj_A^S\,K_A^{\mu}(s)\,\proj_A^S.
    \label{eq:MS_def}
\end{align}
Averaging over Haar-random encoded inputs on the two-dimensional support $S$ (see Appendix~\ref{app:haar_average} for the derivation), and summing over all nine qutrit Bell-measurement outcomes followed by the corresponding Heisenberg--Weyl correction, so that no additional factor $1/9$ enters the heralded metrics, the successful-branch probability, the success-weighted effective fidelity, and the conditional fidelity are
\begin{align}
    \bar p_{\mathrm{succ}}^{(\mathrm{unprot})}(t,s)
    =\;&
    \bigl[\sigma_{\min}^{S}\!\big(K_A^{\mu}(t)\big)\bigr]^{2}
    \sigma_{\min}^2\!\big(K_C^{\chi}(T)\big)
    \nonumber\\
    &\times\,\frac{\tr_S G_S^{\mu}(s)}{m},
    \label{eq:psucc_unprot}
    \\
    \bar F_{\mathrm{eff}}^{(\mathrm{unprot})}(t,s)
    =\;&
    \bigl[\sigma_{\min}^{S}\!\big(K_A^{\mu}(t)\big)\bigr]^{2}
    \sigma_{\min}^2\!\big(K_C^{\chi}(T)\big)
    \nonumber\\
    &\times\,\frac{1}{m(m+1)}\Bigl[
    \tr_S\!\big(M_S^{\mu}(s)M_S^{\mu\dagger}(s)\big)
    \nonumber\\
    &\hspace{5.4em}
    +\big|\tr_S M_S^{\mu}(s)\big|^2\Bigr],
    \label{eq:Feff_unprot}
\end{align}
\begin{align}
    \bar F_{\mathrm{cond,w}}^{(\mathrm{unprot})}
    =\;&
    \frac{1}{(m+1)\,\tr_S G_S^{\mu}(s)}
    \nonumber\\
    &\times\!\left[
    \tr_S\!\big(M_S^{\mu}(s)M_S^{\mu\dagger}(s)\big)
    +\big|\tr_S M_S^{\mu}(s)\big|^2\right].
    \label{eq:Fcond_unprot}
\end{align}
Equations~\eqref{eq:psucc_unprot}--\eqref{eq:Fcond_unprot} are the Architecture-A performance formulas. They show that the residual support dynamics $G_S^{\mu}(s)$ and $M_S^{\mu}(s)$ control the tradeoff. In the limit $s\to 0$ one has $K_A^{\mu}(s)\to I$ and $G_S^{\mu}(s),M_S^{\mu}(s)\to\proj_A^S$, so that the equations reduce to the protected-sender benchmark of Eq.~\eqref{eq:protocol_I_eff} with $\bar F_{\mathrm{cond,w}}=1$. For finite $s$, the unreversed sender map can reduce both success probability and fidelity. This loss is a consequence of the stipulated fixed receiver filter, not a universal limitation of receiver-side recovery.

In Eqs.~\eqref{eq:psucc_unprot}--\eqref{eq:Fcond_unprot}, the symbol $\bar F_{\mathrm{cond,w}}^{(\mathrm{unprot})}=\bar F_{\mathrm{eff}}^{(\mathrm{unprot})}/\bar p_{\mathrm{succ}}^{(\mathrm{unprot})}$ is the \emph{success-weighted} conditional fidelity, not the uniform Haar average of the per-input normalized conditional fidelity. The two need not coincide for finite residual delay $s$; they coincide at $s=0$, and whenever the success weight is input independent. All calculations below impose $H_A=H_C=0$. The same formulas have an interaction-picture interpretation only when the transformation leaves the displayed jump operators time independent.

\subsection{Fidelity-constrained fixed recovery timing in Architecture A}\label{subsec:fixed_timing}

A comparison of two arbitrarily selected intervention times is not a timing optimization. We first consider a fixed classical schedule: one intervention time $\tau$ is chosen before the experiment and used in every run. The word ``fixed'' refers to this timing choice; the PR operation itself remains probabilistic and heralded. We fix the total duration $T$, set $t=\tau$ and $s=T-\tau$, and optimize over $0\leq\tau\leq T$. Consider the representative coherent $V$-type sender, $\Lambda$-type receiver, and $S=\{0,2\}$, with a common rate $\Gamma$. Their zero-Hamiltonian no-jump maps are
\begin{align}
K_A^V(t)&=
\begin{pmatrix}
\frac{1+e^{-\Gamma t}}{2}&\frac{e^{-\Gamma t}-1}{2}&0\\
\frac{e^{-\Gamma t}-1}{2}&\frac{1+e^{-\Gamma t}}{2}&0\\
0&0&1
\end{pmatrix},
\label{eq:KV_explicit}
\\
K_C^\Lambda(t)&=\diag(e^{-\Gamma t},1,1).
\label{eq:KLambda_explicit}
\end{align}
Introduce the dimensionless variables
\begin{equation}
    x=\Gamma T,
    \qquad y=\Gamma\tau,
    \qquad 0\leq y\leq x.
\end{equation}
Since the receiver factor $p_C(T)=\sigma_{\min}^2[K_C^\Lambda(T)]=e^{-2x}$ is fixed, maximizing the total success probability is equivalent to maximizing
\begin{align}
P_x(y)
&\equiv\frac{\bar p_{\mathrm{fix}}(\tau;T)}{p_C(T)}
=\frac{(1+e^{-2y})[3+e^{-2(x-y)}]}{8},
\label{eq:fixed_timing_success}
\\
\mathcal F_x(y)
&\equiv\bar F_{\mathrm{cond,w}}^{(\mathrm{fix})}(\tau;T)
=\frac{e^{-2(x-y)}+4e^{-(x-y)}+7}
{3[e^{-2(x-y)}+3]}.
\label{eq:fixed_timing_fidelity}
\end{align}
The linear-exponential term $4e^{-(x-y)}$ in Eq.~\eqref{eq:fixed_timing_fidelity} is the coherent-overlap contribution and must not be replaced by a squared survival factor.

\begin{proposition}[Fixed-schedule sender timing]\label{prop:fixed_timing}
Within Architecture A and the model specified above, assume that one intervention time is fixed in advance and used in every run. Let $\Gamma>0$ and $T>0$, so $x=\Gamma T>0$, and let $0\leq F_0\leq1$. Consider
\begin{equation}
\max_{0\leq y\leq x}P_x(y)
\quad\text{subject to}\quad
\mathcal F_x(y)\geq F_0.
\label{eq:fixed_timing_optimization}
\end{equation}
Define
\begin{align}
y_c&=\frac{\ln3}{2},
\label{eq:timing_c}
\\
F_{\mathrm{nr}}(x)&=\mathcal F_x(0)
=\frac{e^{-2x}+4e^{-x}+7}{3(e^{-2x}+3)},
\label{eq:F_no_reversal}
\end{align}
When the fidelity constraint is active, $F_0>F_{\mathrm{nr}}(x)$, define
\begin{equation}
z_F=\frac{2-\sqrt{3(1-F_0)(9F_0-1)}}{3F_0-1},
\qquad
y_F=x+\ln z_F.
\label{eq:earliest_feasible_time}
\end{equation}
Here ``nr'' denotes no sender reversal; the receiver operation is still present. If $0<x\leq y_c$, the optimizer is $y_{\mathrm{opt}}=0$ for $F_0\leq F_{\mathrm{nr}}(x)$ and $y_{\mathrm{opt}}=y_F$ for $F_{\mathrm{nr}}(x)<F_0\leq1$; at $F_0=1$, $y_F=x$.

If $x>y_c$, additionally define
\begin{equation}
F_\times(x)=\mathcal F_x(y_c)
=\frac{7+4\sqrt3e^{-x}+3e^{-2x}}{9(1+e^{-2x})}.
\label{eq:F_timing_cross}
\end{equation}
Then
\begin{equation}
\operatorname*{arg\,max}_{\substack{0\leq y\leq x\\ \mathcal F_x(y)\geq F_0}}P_x(y)
=
\begin{cases}
\{0\},&F_0\leq F_{\mathrm{nr}}(x),\\
\{y_F\},&F_{\mathrm{nr}}(x)<F_0<F_\times(x),\\
\{y_c,x\},&F_0=F_\times(x),\\
\{x\},&F_\times(x)<F_0\leq1.
\end{cases}
\label{eq:fixed_timing_regions}
\end{equation}
Thus an intermediate optimum, when present, is the active fidelity boundary $y_F$, not an unconstrained stationary maximum of the success probability.
\end{proposition}

\paragraph*{Proof.}
Let $a=e^{-(x-y)}$. Equation~\eqref{eq:fixed_timing_fidelity} obeys
\begin{equation}
\frac{d\mathcal F_x}{da}
=\frac{4(1-a)(a+3)}{3(a^2+3)^2}\geq0,
\end{equation}
so the feasible set is $[y_F,x]$ when the constraint is active. Direct differentiation gives
\begin{equation}
P_x''(y)=\frac{3e^{-2y}+e^{-2x}e^{2y}}{2}>0,
\end{equation}
so a maximum on that interval must occur at $y_F$ or $x$. Their ordering follows from
\begin{equation}
P_x(x)-P_x(y)
=\frac{[1-e^{-2(x-y)}][1-3e^{-2y}]}{8},
\label{eq:fixed_endpoint_difference}
\end{equation}
whose sign changes at $y=y_c$. Solving $\mathcal F_x(y_F)=F_0$ yields Eq.~\eqref{eq:earliest_feasible_time}, completing the cases above. \hfill$\square$

For the cut used below, $x=2$, the thresholds are
\begin{equation}
F_{\mathrm{nr}}(2)=0.8348648359,
\qquad
F_\times(2)=0.8720912323.
\label{eq:timing_thresholds_T2}
\end{equation}
At the second threshold, $y_F=y_c$ and $y=x=2$ are degenerate optima. This proposition is an optimization over one preassigned, nonrandom timing used in every run, within the prescribed family of maximal exact sender PR operations and the fixed receiver Architecture A.

\begin{corollary}[Pooled-fidelity time sharing]
\label{cor:randomized_timing}
Retain Architecture A and the model of Proposition~\ref{prop:fixed_timing}, but let the intervention time $y$ be drawn before each run from an input-independent probability law $\nu$ on $[0,x]$. Define
\begin{equation}
\begin{aligned}
P_\nu&=\int_0^x P_x(y)\,d\nu(y),&
E_\nu&=\int_0^x E_x(y)\,d\nu(y),\\
\mathcal F_\nu&=\frac{E_\nu}{P_\nu},&
E_x(y)&=P_x(y)\mathcal F_x(y).
\end{aligned}
\label{eq:randomized_metrics}
\end{equation}
The actual success probability is $p_C(T)P_\nu$. If $x\geq y_c=(\ln3)/2$ and only the pooled successful ensemble is required to satisfy $\mathcal F_\nu\geq F_0$, the global optimum over all such timing distributions is
\begin{equation}
\nu_*=\begin{cases}
\delta_0,
&0\leq F_0\leq F_{\mathrm{nr}}(x),\\[1mm]
(1-\lambda_*)\delta_0+\lambda_*\delta_x,
&F_{\mathrm{nr}}(x)<F_0\leq1,
\end{cases}
\label{eq:optimal_randomized_schedule}
\end{equation}
where
\begin{align}
P_{\mathrm{nr}}&=P_x(0)=\frac{3+e^{-2x}}4,&
P_{\mathrm{end}}&=P_x(x)=\frac{1+e^{-2x}}2.
\label{eq:randomized_endpoint_probabilities}
\end{align}
For the second case in Eq.~\eqref{eq:optimal_randomized_schedule}, define $\Delta F_0=F_0-F_{\mathrm{nr}}(x)>0$ and
\begin{equation}
\lambda_*=
\frac{P_{\mathrm{nr}}\Delta F_0}
{P_{\mathrm{nr}}\Delta F_0+P_{\mathrm{end}}(1-F_0)}.
\label{eq:optimal_endpoint_weight}
\end{equation}
At $F_0=1$ this gives $\lambda_*=1$. Here $\delta_0$ denotes omission of sender PR and $\delta_x$ denotes PR immediately before the Bell measurement.
\end{corollary}

\paragraph*{Proof.}
Put $\zeta=e^{-x}$ and $a=e^{-(x-y)}\in[\zeta,1]$, and define the affine function
\begin{equation}
\ell_x(P)=
\frac{1+5\zeta}{3(1+\zeta)}P
+\frac{(1-\zeta)(1+\zeta^2)}{3(1+\zeta)}.
\label{eq:randomized_chord}
\end{equation}
Direct substitution into Eqs.~\eqref{eq:fixed_timing_success} and~\eqref{eq:fixed_timing_fidelity} gives
\begin{align}
\ell_x(P_x(y))-E_x(y)
={}&\frac{(a-\zeta)(a-1)}{6(1+\zeta)a^2}
\nonumber\\
&\times\bigl[\zeta a^2+(\zeta^2-1)a
 +\zeta(2\zeta-1)\bigr].
\label{eq:randomized_chord_factorization}
\end{align}
For $x\geq(\ln3)/2$, one has $\zeta\leq1/\sqrt3$. The bracketed quadratic is convex in $a$ and has endpoint values $2\zeta(\zeta^2+\zeta-1)<0$ and $3\zeta^2-1\leq0$, so it is nonpositive throughout $[\zeta,1]$. Since $(a-\zeta)(a-1)\leq0$,
\begin{equation}
E_x(y)\leq\ell_x(P_x(y)),
\label{eq:randomized_chord_bound}
\end{equation}
with equality only at $y=0$ and $y=x$. Averaging preserves this bound, hence $E_\nu\leq\ell_x(P_\nu)$.

For $F_0\leq F_{\mathrm{nr}}(x)$, $y=0$ is feasible and uniquely maximizes $P_x(y)$ by the strict convexity established in Proposition~\ref{prop:fixed_timing} and $P_{\mathrm{nr}}>P_{\mathrm{end}}$. For $F_0>F_{\mathrm{nr}}(x)$, write $\ell_x(P)=\kappa_xP+\beta_x$. Directly,
\begin{equation}
F_{\mathrm{nr}}(x)-\kappa_x
=\frac{4(1-\zeta)(1+\zeta^2)}
{3(1+\zeta)(3+\zeta^2)}>0.
\end{equation}
The pooled constraint therefore implies
\begin{equation}
F_0P_\nu\leq E_\nu\leq\kappa_xP_\nu+\beta_x,
\qquad
P_\nu\leq\frac{\beta_x}{F_0-\kappa_x}.
\end{equation}
The endpoint distribution in Eq.~\eqref{eq:optimal_randomized_schedule}, with Eq.~\eqref{eq:optimal_endpoint_weight}, satisfies $\mathcal F_{\nu_*}=F_0$ and saturates this bound. It is therefore globally optimal; the strictness of Eq.~\eqref{eq:randomized_chord_bound} away from the endpoints also makes its support unique when the constraint is active. \hfill$\square$

At $x=2$ and $F_0=0.86$, Proposition~\ref{prop:fixed_timing} gives $y_F=0.3953035$ and the actual success probability $p_C(T)P_x(y_F)=1.01180\%$. Corollary~\ref{cor:randomized_timing} instead uses final-time PR with probability $\lambda_*=0.2101581$ and omits sender PR otherwise, giving $p_C(T)P_{\nu_*}=1.28759\%$ at the same pooled fidelity. This is a $27.26\%$ relative increase. If every timing used with nonzero probability must separately satisfy $\mathcal F_x(y)\geq F_0$, however, then $P_\nu\leq\max_{\mathcal F_x(y)\geq F_0}P_x(y)$, and randomization cannot improve on Proposition~\ref{prop:fixed_timing}. The endpoint-mixture result is not asserted for $x<(\ln3)/2$, where the same chord is not a global upper bound.

\subsection{Architecture B: full Bell-branch receiver recovery}\label{subsec:full_branch_receiver}

Architecture B waits for the Bell outcome $(r,n)$ and reverses the complete effective transfer branch rather than first applying a fixed full-space receiver inverse. With the Bell convention of Appendix~\ref{app:qutrit_teleportation}, define the support-restricted transfer map
\begin{equation}
\mathcal T_{r,n}^{\mu,\chi,S}(\tau;T)
=K_C^\chi(T)X_C^rZ_C^{-n}K_A^\mu(T-\tau)Q_S.
\label{eq:full_branch_transfer}
\end{equation}
Let $Q_S^{(C)}:\mathbb C^m\to\mathcal H_C$ denote the receiver-side
copy of the encoding isometry, with
$Q_S^{(C)}Q_S^{(C)\dagger}=\Pi_S^{(C)}$.
The successful maximal exact sender PR at $\tau$ contributes the common amplitude
\begin{equation}
c_A^{\mu,S}(\tau)=\sigma_{\min}^{S}\!\bigl(K_A^\mu(\tau)\bigr),
\end{equation}
and the normalized Bell projection contributes $1/3$. The complete pre-recovery branch from the logical input to $C$ is therefore
\begin{equation}
\mathcal A_{r,n}(\tau;T)
=\frac{c_A^{\mu,S}(\tau)}{3}\,
\mathcal T_{r,n}^{\mu,\chi,S}(\tau;T).
\label{eq:full_branch_A}
\end{equation}
When $\mathcal T_{r,n}$ has full column rank, the optimal exact receiver element for this branch is
\begin{equation}
M_{C|r,n}^{\mathrm{br}}(\tau;T)
=\sigma_{\min}(\mathcal T_{r,n})Q_S^{(C)}\mathcal T_{r,n}^{+},
\label{eq:full_branch_receiver_filter}
\end{equation}
which satisfies
\begin{equation}
M_{C|r,n}^{\mathrm{br}}\mathcal A_{r,n}
=\frac{c_A^{\mu,S}}{3}
\sigma_{\min}(\mathcal T_{r,n})Q_S^{(C)}.
\end{equation}
The filter in Eq.~\eqref{eq:full_branch_receiver_filter} already includes the Bell correction; applying $Z_C^nX_C^{-r}$ again would double-correct the branch. Established measurement-reversal theory~\cite{cheong2012balance,lee2021teleportation} then gives
\begin{equation}
p_{\mathrm{br}}^{(B)}(\tau;T)
=\frac{[c_A^{\mu,S}(\tau)]^2}{9}
\sum_{r=0}^{2}\sum_{n=0}^{2}
\sigma_{\min}^2[\mathcal T_{r,n}^{\mu,\chi,S}(\tau;T)].
\label{eq:full_branch_probability}
\end{equation}
The factor $1/9$ is essential: it is the squared Bell-projection amplitude, and the nine mutually exclusive outcomes are then summed. All full-column-rank successful branches have conditional fidelity one. Equation~\eqref{eq:full_branch_probability} is receiver-optimal conditional on the stipulated maximal exact sender PR at $\tau$; arbitrary sender filters or coherent controls are not optimized.

\begin{proposition}[Sender-PR redundancy]\label{prop:sender_redundancy}
Assume the coherent $V$ sender and $\Lambda$ receiver of Eqs.~\eqref{eq:KV_explicit}--\eqref{eq:KLambda_explicit}, equal positive rates $\Gamma>0$, $S=\{0,2\}$, independent Markovian baths with ideal no-jump monitoring, noiseless $B$, and the fixed qutrit Bell basis. Let
\begin{equation}
\begin{aligned}
q&=e^{-2\Gamma T},&
u&=e^{-2\Gamma\tau},\\
v&=e^{-2\Gamma(T-\tau)},&
uv&=q.
\end{aligned}
\label{eq:quv_def}
\end{equation}
Within the sender-operation family specified above, Eq.~\eqref{eq:full_branch_probability} becomes
\begin{align}
p_{\mathrm{br}}^{(B)}(\tau;T)
&=\frac{1+u}{6}
\left[\frac{(1+q)(1+v)}{2}+q\right]
\label{eq:pbranch_closed_1}
\\
&=\frac{1}{12}\Bigl[1+4q+q^2
 +(1+3q)e^{-2\Gamma\tau}\nonumber\\
&\hspace{5em}{}+q(1+q)e^{2\Gamma\tau}\Bigr].
\label{eq:pbranch_closed_2}
\end{align}
For every finite $T>0$, its maximum over $0\leq\tau\leq T$ is attained uniquely at $\tau=0$.
\end{proposition}

\paragraph*{Proof.}
Writing $a=(1+\sqrt v)/2$ and $b=(\sqrt v-1)/2$, the three shift classes, each threefold degenerate in $n$, have
\begin{equation}\label{eq:branch_singular_values}
\begin{aligned}
\sigma_{\min}^2(\mathcal T_{0,n})&=qa^2+b^2,\\
\sigma_{\min}^2(\mathcal T_{1,n})&=q,\\
\sigma_{\min}^2(\mathcal T_{2,n})&=a^2+qb^2.
\end{aligned}
\end{equation}
Their sum gives Eq.~\eqref{eq:pbranch_closed_1}. With $y=\Gamma\tau$,
\begin{equation}
\frac{d^2p_{\mathrm{br}}^{(B)}}{dy^2}
=\frac{(1+3q)e^{-2y}+q(1+q)e^{2y}}{3}>0,
\end{equation}
so the maximum lies at an endpoint. Finally,
\begin{equation}
p_{\mathrm{br}}^{(B)}(0;T)-p_{\mathrm{br}}^{(B)}(T;T)
=\frac{q(1-q)}{6}>0,
\label{eq:pbranch_endpoint_difference}
\end{equation}
which proves the claim. \hfill$\square$

Here $\tau=0$ means that the sender PR reduces to the identity on the encoded support; it does not remove the sender noise or receiver recovery. Rather, Charlie's branch-dependent filter reverses the complete composite transfer map. Classical randomization of the sender-PR time cannot improve this result: every accepted full-column-rank branch has unit conditional fidelity, while Proposition~\ref{prop:sender_redundancy} gives $p_{\mathrm{br}}^{(B)}(\tau;T)<p_{\mathrm{br}}^{(B)}(0;T)$ for every $\tau>0$. Hence every input-independent timing distribution $\nu$ satisfies
\begin{equation}
p_{{\rm br},\nu}^{(B)}
=\int_0^T p_{\mathrm{br}}^{(B)}(\tau;T)\,d\nu(\tau)
\leq p_{\mathrm{br}}^{(B)}(0;T),
\label{eq:branch_randomization_bound}
\end{equation}
with equality only when all weight is at $\tau=0$. At $\Gamma T=2$, Eq.~\eqref{eq:pbranch_closed_1} gives $17.8933\%$, $11.2849\%$, and $17.5936\%$ for $\Gamma\tau=0,1,2$, respectively, all with unit conditional fidelity. In the protected limit, Eq.~\eqref{eq:receiver_outcome_adaptive} gives $p_{C,\mathrm{adapt}}=(1+2e^{-4})/3=0.345544$, compared with the fixed full-space factor $e^{-4}=0.0183156$, a receiver-factor ratio of $18.87$. This ratio compares the two stipulated receiver capabilities in this model; it is not an improvement over the outcome-adaptive protocols already known in Refs.~\cite{lee2021teleportation,im2021optimal}.

A related but distinct branch-selective redundancy occurs in the two-way qubit repeater of Ref.~\cite{jeong2026recovery}: its pre-Bell reversal leaves concurrence unchanged for the $\Psi^\pm$ outcomes because the Bell projection removes the targeted $|11\rangle$ component. Proposition~\ref{prop:sender_redundancy} instead sums all nine qutrit Bell outcomes, uses an outcome-dependent receiver inverse of each complete transfer branch, and optimizes a continuous sender-reversal time. Its new content is the closed all-$T>0$ result within that stated model and operation family, not the first observation of reversal redundancy in any Bell-measurement architecture.

Architecture B does not require a new classical Bell message: standard teleportation already communicates $(r,n)$. Its additional control requirement is a calibrated, switchable family of nonunitary success filters and the corresponding heralding. In the representative model, $K_C^\Lambda$ commutes with $Z$ and $X^rZ^{-n}=\omega^{rn}Z^{-n}X^r$, so
\begin{equation}
\mathcal T_{r,n}(\tau;T)=\omega^{rn}Z^{-n}\mathcal T_{r,0}(\tau;T).
\label{eq:branch_phase_relation}
\end{equation}
Consequently,
$M_{C|r,n}^{\mathrm{br}}=\omega^{-rn}M_{C|r,0}^{\mathrm{br}}Z^n$:
outcomes with the same $r$ share one nonunitary filter core and differ
only by a known phase unitary applied before that core. Thus three
switchable nonunitary settings suffice rather than nine unrelated
filters. Architecture A uses only one fixed nonunitary receiver filter.
The fixed-schedule timing regions of Proposition~\ref{prop:fixed_timing},
together with the pooled-fidelity optimum of
Corollary~\ref{cor:randomized_timing}, quantify the performance
consequence of that receiver-control restriction under the two distinct
scheduling assumptions.

\begin{figure*}[tp]
    \centering
    \includegraphics[width=\textwidth]{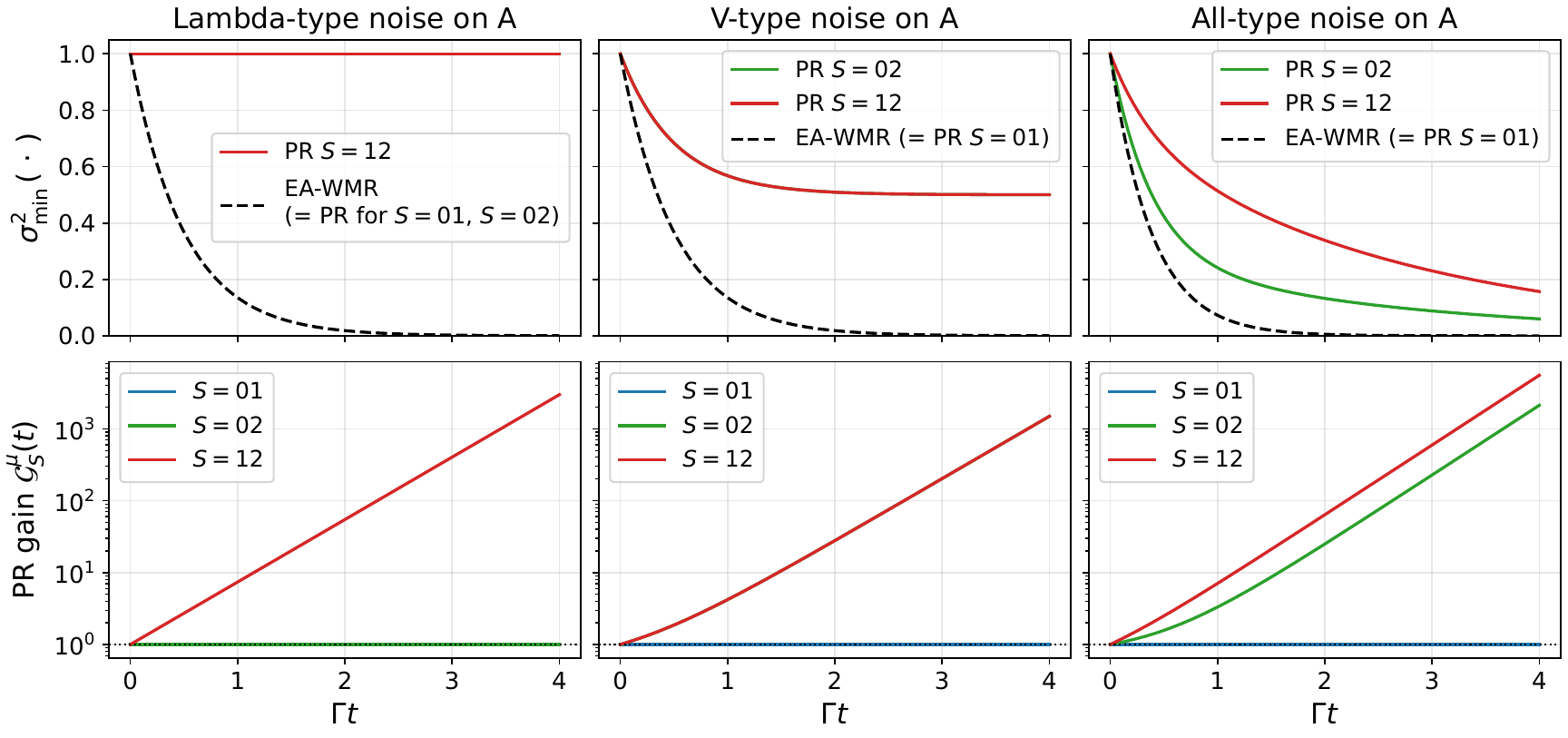}
    \caption{Sender-side support advantage. Top row: $[\sigma_{\min}^{S}(K_0^{\mu}(t))]^{2}=\sigma_{\min}^{2}(K_0^{\mu}(t)Q_S)$ vs. $\Gamma t$ for the three local single-qutrit noise topologies $\mu\in\{\Lambda,V,\mathrm{all}\}$ (columns), with two-level supports $S\in\{01,02,12\}$ (solid colored) and the full-space EA-WMR factor $\sigma_{\min}^{2}(K_0^{\mu}(t))$ (dashed black). Bottom row: PR gain $\mathcal{G}_S^\mu(t)=[\sigma_{\min}^{S}(K_0^{\mu}(t))]^{2}/\sigma_{\min}^{2}(K_0^{\mu}(t))$ on a log scale; $\mathcal G>1$ quantifies the strict PR advantage. The sender Hamiltonian is set to zero. Large gains at long times must be read together with the small absolute success factors in the top row.}
    \label{fig:sender_support_advantage}
\end{figure*}

\begin{figure*}[tp]
    \centering
    \includegraphics[width=\textwidth]{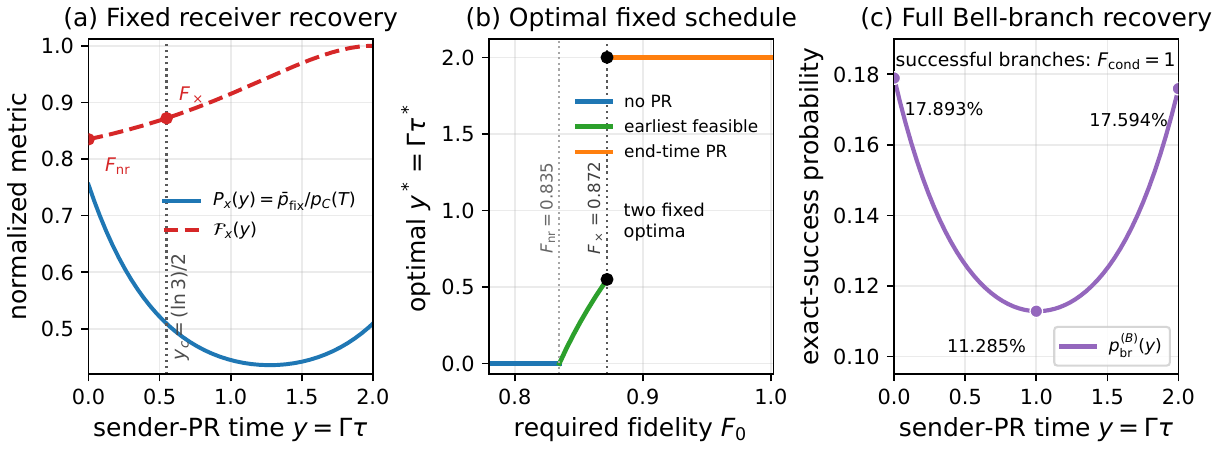}
    \caption{Receiver capability changes the sender-timing problem for a coherent $V$-type sender, a $\Lambda$-type receiver, $S=\{0,2\}$, and $\Gamma T=2$. (a) Architecture A: normalized success probability $P_x(y)=\bar p_{\mathrm{fix}}/p_C(T)$ and success-weighted conditional fidelity $\mathcal F_x(y)$ versus $y=\Gamma\tau$. (b) Optimal single preassigned value of $y$ at a required fidelity $F_0$; randomized pooled-fidelity schedules are not represented by this curve. The vertical boundaries are $F_{\mathrm{nr}}=0.834865$ and $F_\times=0.872091$; at $F_\times$, $y=y_c=(\ln3)/2$ and $y=2$ are degenerate. (c) Architecture B: exact-teleportation success probability from Eq.~\eqref{eq:pbranch_closed_1}. The values at $y=0,1,2$ are $17.8933\%$, $11.2849\%$, and $17.5936\%$, respectively, and every accepted full-column-rank branch has conditional fidelity one.}
    \label{fig:recovery_placement}
\end{figure*}

We emphasize four common boundaries. First, a classically fixed timing schedule does not make the PR operation deterministic: both architectures retain heralded probabilistic filters. The randomized result additionally assumes that the timing draw is independent of the unknown input and that the fidelity floor applies only to the pooled successful ensemble. Second, lack of exact recovery for finite residual sender time applies to Architecture A; Architecture B is exact on successful full-column-rank branches. Third, the local-noise model places the sender and receiver in independent baths and takes $B$ as ideal; a collective sender--partner bath requires a joint encoded-support map (Appendix~\ref{app:block_extension}). Fourth, the timing results use Markovian semigroup composition across the intermediate operation. A non-Markovian bath requires a multi-time description, such as a process tensor~\cite{pollock2018nonmarkovian}, and the whole-interval OU example below is not a timing optimization.

The recovery constructions are not specific to the $V/\Lambda$ pairing or to $S=\{0,2\}$. For another encoded support, its isometry replaces $Q_S$ in the restricted map. The support/full-space comparison then depends on the orientation of that support relative to the least-reversible right-singular subspace, rather than on the level labels. A strict success advantage is therefore possible for other supports, but is not guaranteed merely by using a multilevel carrier. Figure~\ref{fig:sender_support_advantage} illustrates this distinction for three qutrit decay topologies and all three two-level computational supports.

Likewise, under the same independent local Markovian noise and ideal monitoring assumptions, unequal decay rates or a nonzero system Hamiltonian enter through the corresponding no-jump maps in Eq.~\eqref{eq:full_branch_transfer}. A full-column-rank transfer branch still admits exact support recovery by Eq.~\eqref{eq:full_branch_receiver_filter}, and Eq.~\eqref{eq:full_branch_probability} gives the corresponding total success probability after summing over Bell outcomes. The optimal sender time, however, depends on the resulting singular values. The convexity arguments, timing thresholds, and three-setting simplification above are properties of the specified coherent $V/\Lambda$ model, not general consequences of branch invertibility. Thus the framework provides a way to compare receiver capabilities in other models without presupposing that the same recovery placement will remain optimal.

Figure~\ref{fig:recovery_placement} summarizes the timing results for the representative model. The randomized schedule is a distribution rather than a single plotted value of $y$.

\subsection{Supporting mechanism: collective two-qubit Markovian and OU branches}\label{subsec:two_qubit_supporting}

The qutrit results above were derived under the local-noise factorization $K_A^\mu(t)\otimes I_B\otimes K_C^\chi(t+s)$. To check that the support-restricted singular-value gap of Eq.~\eqref{eq:pr_geq_ea} is not tied to diagonal qutrit models, we summarize as a supporting mechanism the closed-form result for a collective two-qubit dissipative branch in both the Markovian and non-Markovian Ornstein--Uhlenbeck (OU) regimes. The closed-form expressions follow from the bright/dark decomposition of the collective no-jump branch; only the final results are needed in the qutrit teleportation context, so we present them directly.

Take the ordered basis $\{|11\rangle,|10\rangle,|01\rangle,|00\rangle\}$ with collective system operators
\begin{equation}\label{eq:two_qubit_hamiltonian}
    H_{\mathrm{sys}}
    =
    \frac{\omega}{2}\left(\sigma_z\otimes \id + \id\otimes \sigma_z\right),
    \qquad
    L
    =
    \sigma_- \otimes \id + \id \otimes \sigma_- ,
\end{equation}
and an encoded input on the single-excitation/ground sector,
\begin{equation}\label{eq:two_qubit_initial_state}
    |\psi_S(0)\rangle=\beta |10\rangle+\alpha |00\rangle,
    \qquad
    \proj = \projket{10}+\projket{00}.
\end{equation}
Let $|s\rangle=(|10\rangle+|01\rangle)/\sqrt{2}$ and $|a\rangle=(|10\rangle-|01\rangle)/\sqrt{2}$ be the bright and dark single-excitation modes, and let $G(t)$ be the no-jump amplitude of $|s\rangle$. The dark and ground amplitudes have unit modulus. Thus, on the zero/single-excitation sector
\begin{equation}\label{eq:two_qubit_low_excitation_sector}
    \mathcal H_{\leq 1}=\operatorname{span}\{|00\rangle,|10\rangle,|01\rangle\},
\end{equation}
the no-jump map has singular values $\{1,1,|G(t)|\}$, whereas its restriction to the encoded support in Eq.~\eqref{eq:two_qubit_initial_state} has singular values $\{1,\sqrt{[1+|G(t)|^2]/2}\}$. A direct calculation therefore gives
\begin{equation}\label{eq:two_qubit_pr_vs_ea}
    p_{\mathrm{PR}}^{(S)}(t)=\frac{1+|G(t)|^2}{2},
    \qquad
    p_{\mathrm{EA}}^{(\leq 1)}(t)=|G(t)|^2.
\end{equation}
For $G(t)\neq0$, the second quantity is the EA-WMR benchmark on $\mathcal H_{\leq1}$; at isolated zeros of $G(t)$, exact sector-wide inversion is undefined and $|G(t)|^2=0$ denotes only the continuous smallest-singular-value scale plotted below. In either case it is not automatically the benchmark on the full ordered four-dimensional space. To display the missing sector explicitly, define the double-excitation vacuum-to-vacuum amplitude
\begin{equation}\label{eq:G2_definition}
    G_2(t)=\langle 11|\langle 0_E|U(t)|11\rangle|0_E\rangle.
\end{equation}
Excitation-number conservation then yields, whenever the full no-jump map is invertible,
\begin{equation}\label{eq:two_qubit_full_ea}
\begin{aligned}
    p_{\mathrm{EA}}^{(4)}(t)
    &=\sigma_{\min}^2\!\bigl(K_0(t)\bigr)\\
    &=\min\{1,|G(t)|^2,|G_2(t)|^2\}\\
    &\leq p_{\mathrm{EA}}^{(\leq1)}(t).
\end{aligned}
\end{equation}
If $G(t)=0$, exact EA-WMR is already undefined on $\mathcal H_{\leq1}$; if $G_2(t)=0$, exact EA-WMR is undefined on the full four-dimensional space. In the Markovian limit generated by the collective $L$ in Eq.~\eqref{eq:two_qubit_hamiltonian}, $|G_2(t)|=|G(t)|=e^{-\Gamma t}$ (free Hamiltonian phases do not affect the singular values), so the sector and full-space benchmarks coincide and $p_{\mathrm{PR}}^{(S)}-p_{\mathrm{EA}}^{(4)}=(1-e^{-2\Gamma t})/2>0$ for $t>0$. For an OU bath with correlation $\alpha(t,s)=(\Gamma\gamma/2)e^{-\gamma|t-s|}$, whose causal branch is $e^{-\gamma(t-s)}$ for $t\geq s$, the symmetric-mode amplitude obeys
\begin{equation}\label{eq:G_ode}
    \ddot G(t)+\gamma\dot G(t)+\Gamma\gamma G(t)=0,
    \qquad G(0)=1,\ \dot G(0)=0,
\end{equation}
with closed-form solution~\cite{zhao2011dynamics}
\begin{equation}\label{eq:G_solution}
\begin{aligned}
    G(t)&=e^{-\gamma t/2}\!\left[
    \cosh\!\left(\frac{\delta_{\mathrm{OU}}t}{2}\right)
    +\frac{\gamma}{\delta_{\mathrm{OU}}}
    \sinh\!\left(\frac{\delta_{\mathrm{OU}}t}{2}\right)\right],\\
    \delta_{\mathrm{OU}}&=\sqrt{\gamma^2-4\Gamma\gamma}.
\end{aligned}
\end{equation}
Equation~\eqref{eq:G_solution} determines both sector quantities in Eq.~\eqref{eq:two_qubit_pr_vs_ea}, so their gap is controlled by the memory-modified single-excitation amplitude $G(t)$. The exact full-space benchmark in Eq.~\eqref{eq:two_qubit_full_ea}, however, additionally requires a separate calculation of $G_2(t)$. A representative sector comparison across three memory regimes $\gamma/\Gamma\in\{10,2,0.5\}$ is given in Fig.~\ref{fig:ou_twoqubit_compare} of Appendix~\ref{app:OU_process_tensor}.

We emphasize that Eq.~\eqref{eq:two_qubit_pr_vs_ea} describes a single uninterrupted dissipative interval and serves as a supporting mechanism rather than a qutrit teleportation protocol-level result. The OU propagator $G(t)$ does not satisfy a semigroup composition law $G(t+s)=G(s)G(t)$, so this whole-interval analysis cannot be applied directly to configurations in which a PR step is inserted at an intermediate time. A consistent treatment of intermediate intervention in the non-Markovian regime must retain multi-time system--bath correlations, for example through a process tensor~\cite{pollock2018nonmarkovian}; see Appendix~\ref{app:OU_process_tensor} for the corresponding caveat. Extensions to genuine three-level non-Markovian dynamics fit naturally within the time-local quantum-state-diffusion framework~\cite{jing2010nonmarkovian,jing2012time} and are deferred to future work.

Additional protected-sender and topology/support diagnostics are retained in Appendix~\ref{app:jump_operators}.

\subsection{Sender-side support advantage}\label{subsec:numerical_support_advantage}

Figure~\ref{fig:sender_support_advantage} illustrates the spectral criterion in Eq.~\eqref{eq:gain_ratio_exact} for the three local qutrit topologies and two-level supports defined in Appendix~\ref{app:jump_operators}. The top row compares the support-restricted PR and full-space EA-WMR success factors; the bottom row shows their ratio.

For the $V$-type and all-transition models, the bright eigenmodes of $L^{\mu\dagger}L^\mu$ are not computational-basis vectors. For $t>0$, strict gain occurs exactly when the support has trivial intersection with the most strongly decaying eigenspace. For the $\Lambda$-type model, $L^{\Lambda\dagger}L^\Lambda=2|0\rangle\langle0|$, so supports containing $|0\rangle$ have unit gain. The orthogonal support $S=\{1,2\}$ has gain $\mathcal G=1/\sigma_{\min}^2(K_0^\Lambda(t))$, which increases without bound as the EA-WMR factor decays. Thus a large ratio at long times must be read together with the absolute success factor.

\subsection{Protected-sender benchmark}\label{subsec:numerical_protected}

The protected-sender effective fidelity of Eq.~\eqref{eq:protocol_I_eff} factorizes as
\[
    \bar F_{\mathrm{eff}}^{(\mathrm{prot})}(t,s)
    =
    \bigl[\sigma_{\min}^{S}\!\big(K_A^{\mu}(t)\big)\bigr]^{2}\,
    \sigma_{\min}^{2}\!\big(K_C^{\chi}(t+s)\big).
\]
The first factor is the support-restricted sender factor shown in Fig.~\ref{fig:sender_support_advantage}; the second is the full-space receiver factor over $T=t+s$. The corresponding zoomed heatmap is reported in Appendix~\ref{app:jump_operators}, Fig.~\ref{fig:supp_sender_diagnostics}(a). It is a diagnostic for Architecture A rather than a comparison with full Bell-branch recovery.

\subsection{Recovery placement under the two receiver capabilities}\label{subsec:numerical_recovery_placement}

Figure~\ref{fig:recovery_placement} evaluates Propositions~\ref{prop:fixed_timing} and~\ref{prop:sender_redundancy} at $\Gamma T=2$. Panel~(a) shows why maximizing the Architecture-A success probability alone is insufficient: omitting sender PR maximizes $P_x(y)$, but the conditional fidelity increases with the intervention time. Panel~(b) plots the fixed-schedule constrained optimizer. The intermediate branch is not a stationary maximum; it is the earliest intervention that satisfies $F_0$. The jump at $F_\times$ reflects the exchange of the two endpoint maxima, with both $y=y_c$ and $y=x$ optimal exactly at the threshold. The randomized optimizer of Corollary~\ref{cor:randomized_timing} is not represented by one value of $y$: it is a classical distribution over the two endpoints. Panel~(c) shows the different objective under Architecture B. All successful full-column-rank branches now have unit conditional fidelity, and the exact-teleportation probability is maximal at $\tau=0$ for the whole interval.

\section{Conclusion}\label{sec:conclusion}

We have used a support-restricted no-jump inverse, termed Partial Reversal (PR), to determine when an intermediate sender recovery is useful in noisy encoded-qutrit teleportation. The pseudoinverse and smallest-singular-value reversal bound are established tools; the protocol result here is the dependence of optimal recovery placement on both the receiver's allowed operation class and the operational meaning of the fidelity constraint.

At the operator level, PR is the Moore--Penrose left inverse of the restricted no-jump map $B(t)=K_0(t)Q$, scaled by $\sigma_{\min}(B)$. Its maximal exact-reversal probability is $p_{\mathrm{PR}}^{\max}(t)=\sigma_{\min}^{2}(B(t))$. The comparison with full-space EA-WMR is governed by the spectral criterion of Eq.~\eqref{eq:gain_ratio_exact}: the two probabilities are equal exactly when the encoded support intersects the full-space minimum right-singular subspace $E_{\min}(t)$, and the restricted probability is strictly larger when that intersection is trivial. The restricted inverse can also remain well defined when $K_0(t)$ is singular on an unused carrier direction, provided $K_0(t)Q$ retains full column rank.

For Architecture A, in which Charlie applies one outcome-independent full-space filter followed by the usual Bell-dependent unitary, residual sender evolution remains after an intermediate PR. The resulting Haar-averaged success probability, success-weighted fidelity, and success-weighted conditional fidelity are given by Eqs.~\eqref{eq:psucc_unprot}--\eqref{eq:Fcond_unprot}. For the coherent $V$ sender, $\Lambda$ receiver, and $S=\{0,2\}$, Proposition~\ref{prop:fixed_timing} solves the fixed-$T$ problem among single preassigned recovery times: depending on $T$ and $F_0$, the optimizer is no sender PR, the earliest feasible intermediate PR, or a PR immediately before the Bell measurement. At $\Gamma T=2$, the regime boundaries are $F_{\mathrm{nr}}=0.8348648359$ and $F_\times=0.8720912323$, with a two-point degeneracy at the latter value. If the same fidelity floor must hold separately for every timing in use, classical randomization cannot improve this result. Under the distinct requirement that only the pooled successful ensemble meet the floor, Corollary~\ref{cor:randomized_timing} proves that, for $\Gamma T\geq(\ln3)/2$, the global optimum over all input-independent timing distributions time-shares between no sender PR and final-time PR. At $\Gamma T=2$ and $F_0=0.86$, this raises the total success probability from $1.01180\%$ for the best fixed schedule to $1.28759\%$.

Architecture B allows Charlie to condition a nonunitary filter on the Bell result and to invert the complete transfer branch. Every accepted full-column-rank branch then has unit conditional fidelity. Proposition~\ref{prop:sender_redundancy} shows that, within the prescribed family of maximal exact sender PR operations and for the same representative model, the exact-teleportation success probability is maximized uniquely at $\tau=0$ for every finite $T>0$; averaging over recovery times cannot improve a pointwise unique maximum. Thus the sender PR is redundant in this operation class: its reversal task is more efficiently absorbed into the receiver's composite branch filter. This is a model- and operation-class result, not a global optimum over arbitrary sender filters, encoding rotations, coherent controls, or receiver instruments.

The pathway-coherence calculation further separates relative gain from absolute usefulness. For $S=\{0,2\}$, $p_{\mathrm{PR}}=e^{-\Gamma t}\cosh(\eta\Gamma t)$: a dark-state plateau survives only at full coherence $\eta=1$, whereas every fixed $\eta<1$ gives vanishing long-time success even though the PR/EA ratio may grow. Finally, the collective two-qubit Ornstein--Uhlenbeck example supports the same whole-interval singular-value mechanism but is not a non-Markovian intermediate-control protocol. Such a protocol requires a multi-time process description. Natural extensions are therefore to optimize over broader sender and receiver instruments, quantify imperfect monitoring and filter-calibration costs, and treat non-Markovian interventions with an explicit process tensor.

\appendix

\section{Support-restricted pseudoinverse PR construction}\label{app:pr_pseudoinverse_construction}

This appendix collects the canonical support-restricted pseudoinverse construction used throughout the main text and casts it in a form that is convenient for explicit numerical implementation.

Let $\mathcal{H}$ be a finite-dimensional Hilbert space with $\dim\mathcal{H}=d$, let $\Pi=\Pi^\dagger=\Pi^2$ be the orthogonal projector onto the encoded support $\mathcal{H}_{\mathrm{enc}}\subseteq\mathcal{H}$, and let $m=\rank(\Pi)$. Choose any isometry $Q:\mathbb{C}^m\to\mathcal{H}$ whose columns form an orthonormal basis of $\mathcal{H}_{\mathrm{enc}}$, so that $Q^\dagger Q=I_m$ and $QQ^\dagger=\Pi$. Given the no-jump Kraus operator $K_0(t)$, define the restricted matrix
\begin{equation}\label{eq:appA_B_def}
    B(t)=K_0(t)Q,
    \qquad
    B(t)\in\mathcal{L}(\mathbb{C}^m,\mathcal{H}).
\end{equation}
The thin singular-value decomposition of $B(t)$ is $B(t)=U(t)\Sigma(t)V^\dagger(t)$, where $\Sigma(t)=\diag(\sigma_1(t),\ldots,\sigma_m(t))$ with $\sigma_1(t)\ge\cdots\ge\sigma_m(t)\ge 0$. Equivalently, $\sigma_i(t)$ are the singular values of the restricted map $K_0(t)|_{\mathcal{H}_{\mathrm{enc}}}:\mathcal{H}_{\mathrm{enc}}\to\mathcal{H}$, represented in the chosen support basis by $B(t)=K_0(t)Q$. We deliberately avoid the notation $\sigma_{\min}(K_0(t)\Pi)$ for the full-space operator $K_0(t)\Pi$, whose smallest singular value is $0$ whenever $\Pi\neq I$; the support-restricted singular values are those of $B(t)$.

Provided $\sigma_m(t)>0$, equivalently $\rank B(t)=m$, the canonical support-restricted left inverse is the Moore--Penrose pseudoinverse of the restricted map,
\begin{align}\label{eq:appA_RPR_canonical}
    R_{\mathrm{PR}}(t)
    &=
    Q\,B^{+}(t)
    =
    Q\bigl[B^\dagger(t)B(t)\bigr]^{-1}B^\dagger(t)
    \nonumber\\
    &=
    Q\,V(t)\Sigma^{-1}(t)U^\dagger(t).
\end{align}
A direct calculation using $Q^\dagger Q=I_m$ and $QQ^\dagger=\Pi$ verifies the defining left-inverse property
\begin{align}\label{eq:appA_left_inverse_verify}
    R_{\mathrm{PR}}(t)K_0(t)\Pi
    &=
    Q\bigl[B^\dagger(t)B(t)\bigr]^{-1}B^\dagger(t)B(t)Q^\dagger
    \nonumber\\
    &=QQ^\dagger=\Pi.
\end{align}
The minimal operator norm is attained by Eq.~\eqref{eq:appA_RPR_canonical}, namely $\|R_{\mathrm{PR}}(t)\|_{\mathrm{op}}=1/\sigma_m(t)$, and the maximal PR success probability is
\begin{equation}\label{eq:appA_pPR_max}
    p_{\mathrm{PR}}^{\max}(t)
    =
    \frac{1}{\|R_{\mathrm{PR}}(t)\|_{\mathrm{op}}^{2}}
    =
    \sigma_{\min}^{2}(B(t))
    =
    \sigma_m^{2}(t).
\end{equation}
Every exact restricted left inverse whose output is constrained to $\mathcal{H}_{\mathrm{enc}}$ has the form
\begin{equation}\label{eq:appA_general_left_inverse}
    R(t)=Q\,B^{+}(t)+Q\,Z(t)\bigl[I-B(t)B^{+}(t)\bigr],
\end{equation}
where $Z(t)\in\mathcal{L}(\mathcal{H},\mathbb{C}^{m})$ is arbitrary. The dimensions of the building blocks are $B(t):\mathbb{C}^{m}\to\mathcal{H}$, $B^{+}(t):\mathcal{H}\to\mathbb{C}^{m}$, and $I-B(t)B^{+}(t):\mathcal{H}\to\mathcal{H}$, so that the second term acts only on the orthogonal complement of $\operatorname{Ran}B(t)$ and lands back in $\mathcal{H}_{\mathrm{enc}}$. The choice $Z(t)=0$ recovers the Moore--Penrose pseudoinverse and is the canonical minimum-norm choice used in the main text; it also attains the minimal possible operator norm $1/\sigma_{\min}(B(t))$, equal to the bound that controls the PR scaling factor and therefore the maximal success probability. Other choices of $Z(t)$ act nontrivially only outside $\operatorname{Ran}B(t)$ and are not used in this paper; they may increase the physical measurement norm and hence the required POVM scaling, but cannot improve the maximal success probability beyond Eq.~\eqref{eq:appA_pPR_max}. Hand-picked principal-submatrix inverses and projected full-space inverses such as $\Pi K_0^{-1}(t)$ are particular cases of Eq.~\eqref{eq:appA_general_left_inverse} corresponding to specific nonzero $Z(t)$.

In numerical implementations, one should not attempt to invert the full $d\times d$ matrix $\Pi K_0^\dagger(t)K_0(t)\Pi$ directly, because it is structurally singular on $(I-\Pi)\mathcal{H}$ whenever $\Pi\neq I$. The stable implementation is to compute the thin SVD of $B(t)$ in Eq.~\eqref{eq:appA_B_def} and threshold the smallest singular value at a tolerance $\varepsilon_{\mathrm{svd}}=\max(d,m)\,\epsilon_{\mathrm{mach}}\,\sigma_{\max}(B(t))$. If $\sigma_m(t)\le\varepsilon_{\mathrm{svd}}$, exact PR on the chosen support is either impossible or numerically unstable; otherwise, Eq.~\eqref{eq:appA_RPR_canonical} returns the canonical inverse and Eq.~\eqref{eq:appA_pPR_max} returns the maximal success probability. For near-singular cases, the Tikhonov-regularized variant
\begin{equation}\label{eq:appA_tikhonov}
    R_{\lambda}(t)=Q\big(B^\dagger(t)B(t)+\lambda I_m\big)^{-1}B^\dagger(t)
\end{equation}
provides a stable diagnostic, but it is not an exact left inverse and should not be confused with the canonical PR operator.

\section{Haar-Average Derivation for the Fixed-Receiver Architecture}\label{app:haar_average}

This appendix derives the Haar-averaged formulas in Eqs.~\eqref{eq:psucc_unprot}--\eqref{eq:Fcond_unprot} for Architecture A with an unprotected sender and one outcome-independent receiver filter. Let $S\subseteq\{0,\ldots,d-1\}$ index the encoded support of dimension $m=|S|$ on Alice's qudit, and let $\proj_A^S$ denote the corresponding orthogonal projector. We parameterize the encoded input as
\begin{equation}
    |\psi_S\rangle_A=\sum_{j\in S}a_j|j\rangle_A,
    \qquad
    \sum_{j\in S}|a_j|^2=1,
\end{equation}
with $(a_j)_{j\in S}$ drawn from the Haar measure on the unit sphere $\mathbb S^{2m-1}\subset\mathbb C^m$. The standard moments of the Haar measure on $\mathbb S^{2m-1}$ are
\begin{align}
    \langle a_j a_k^*\rangle_{\mathrm{Haar}}
    &=\frac{\delta_{jk}}{m},
    \label{eq:haar_first_moment}
    \\
    \langle a_j a_k a_l^* a_p^*\rangle_{\mathrm{Haar}}
    &=\frac{\delta_{jl}\delta_{kp}+\delta_{jp}\delta_{kl}}{m(m+1)}.
    \label{eq:haar_second_moment}
\end{align}
Equivalently, in operator form,
\begin{align}
    \big\langle |\psi_S\rangle\!\langle\psi_S|\big\rangle_{\mathrm{Haar}}
    &=
    \frac{\proj_A^S}{m},
    \label{eq:haar_unit_op_first}
    \\
    \big\langle |\psi_S\rangle\!\langle\psi_S|^{\otimes 2}\big\rangle_{\mathrm{Haar}}
    &=
    \frac{\proj_A^S\otimes\proj_A^S+\mathrm{SWAP}_S(\proj_A^S\otimes\proj_A^S)}{m(m+1)},
    \label{eq:haar_unit_op_second}
\end{align}
where $\mathrm{SWAP}_S$ exchanges the two factors of the support tensor product.

For the unprotected configuration the unnormalized post-PR sender state is
\begin{equation}
    \tilde\rho_A^{\mathrm{out}}(t,s)
    =
    [c_A^{\mu,S}(t)]^2
    K_A^{\mu}(s)\,\proj_A^S\,|\psi_S\rangle\!\langle\psi_S|\,\proj_A^S\,K_A^{\mu\dagger}(s),
\end{equation}
where the leading factor is
\begin{equation}\label{eq:appB_cA_def}
    [c_A^{\mu,S}(t)]^{2}
    =
    \bigl[\sigma_{\min}^{S}\!\big(K_A^{\mu}(t)\big)\bigr]^{2}
    =
    \sigma_{\min}^{2}\!\big(K_A^{\mu}(t)Q_S\big),
\end{equation}
in which the final expression emphasizes that the singular value is evaluated on the restricted encoded domain $\mathcal{H}_S$ via the isometry $Q_S$, not on the full operator $K_A^{\mu}(t)\proj_A^S$ acting on $\mathcal{H}$. The factor comes from the support-restricted PR identity~\eqref{eq:Aside_PR_identity}. The receiver contribution multiplies this by $[c_C^{\chi}(T)]^2=\sigma_{\min}^2(K_C^{\chi}(T))$ via the EA-WMR identity~\eqref{eq:Cside_EA_identity}, with the operation order (no-jump evolution, then EA-WMR, then Heisenberg--Weyl correction once Alice's outcome is known) as discussed in Section~\ref{subsec:architecture_A_protected}. The probabilities below are understood after summing over all nine qutrit Bell-measurement outcomes; since every outcome is accepted and followed by the corresponding Heisenberg--Weyl correction, no additional factor $1/9$ enters the heralded metrics. The two support-restricted residual operators are
\begin{equation}\label{eq:appB_GS_MS_def}
\begin{aligned}
    G_S^{\mu}(s)
    &=\proj_A^S\,K_A^{\mu\dagger}(s)K_A^{\mu}(s)\,\proj_A^S,\\
    M_S^{\mu}(s)
    &=\proj_A^S\,K_A^{\mu}(s)\,\proj_A^S.
\end{aligned}
\end{equation}
in agreement with Eqs.~\eqref{eq:GS_def}--\eqref{eq:MS_def}. In general $G_S^{\mu}(s)\neq M_S^{\mu\dagger}(s)M_S^{\mu}(s)$, because $K_A^{\mu}(s)$ may leak amplitude outside the encoded support so that $K_A^{\mu}(s)\proj_A^S\neq\proj_A^S K_A^{\mu}(s)\proj_A^S$; the sandwich identity $\proj_A^S K^\dagger K\proj_A^S=(\proj_A^S K\proj_A^S)^\dagger(\proj_A^S K\proj_A^S)$ requires inserting an additional projector that is not present here. The operator $G_S^{\mu}$ controls the total successful-branch norm, while $M_S^{\mu}$ controls the coherent overlap with the original encoded state. With these definitions, the per-input success probability is
\begin{align}
    p_{\mathrm{succ}}(\psi_S,t,s)
    =\;&
    \bigl[\sigma_{\min}^{S}\!\big(K_A^{\mu}(t)\big)\bigr]^{2}
    \sigma_{\min}^2\!\big(K_C^{\chi}(T)\big)
    \nonumber\\
    &\times\,
    \langle\psi_S|G_S^{\mu}(s)|\psi_S\rangle.
\end{align}
Averaging over the Haar measure using Eq.~\eqref{eq:haar_first_moment} gives
\begin{align}
    \bar p_{\mathrm{succ}}^{(\mathrm{unprot})}(t,s)
    =\;&
    \bigl[\sigma_{\min}^{S}\!\big(K_A^{\mu}(t)\big)\bigr]^{2}
    \sigma_{\min}^2\!\big(K_C^{\chi}(T)\big)
    \nonumber\\
    &\times\,
    \frac{\tr_S G_S^{\mu}(s)}{m},
\end{align}
which is Eq.~\eqref{eq:psucc_unprot}.

The per-input conditional fidelity, after the ideal Bell measurement on $A$ and the unitary correction on $C$, is the overlap of the residual sender state with the input,
\begin{equation}
    F_{\mathrm{cond}}(\psi_S,s)
    =
    \frac{|\langle\psi_S|M_S^{\mu}(s)|\psi_S\rangle|^2}{\langle\psi_S|G_S^{\mu}(s)|\psi_S\rangle}.
\end{equation}
The success-weighted effective fidelity is then
\begin{align}
    F_{\mathrm{eff}}(\psi_S,t,s)
    &=
    p_{\mathrm{succ}}(\psi_S,t,s)\,F_{\mathrm{cond}}(\psi_S,s)
    \nonumber\\
    &=
    \bigl[\sigma_{\min}^{S}\!\big(K_A^{\mu}(t)\big)\bigr]^{2}
    \sigma_{\min}^2\!\big(K_C^{\chi}(T)\big)
    \nonumber\\
    &\quad\times
    \bigl|\langle\psi_S|M_S^{\mu}(s)|\psi_S\rangle\bigr|^{2}.
\end{align}
Using Eq.~\eqref{eq:haar_second_moment} to evaluate $\langle |\langle\psi_S|M_S^{\mu}(s)|\psi_S\rangle|^2\rangle_{\mathrm{Haar}}$ gives
\begin{equation}
    \big\langle|\langle\psi_S|M_S^{\mu}|\psi_S\rangle|^2\big\rangle_{\mathrm{Haar}}
    =
    \frac{\tr_S\!\big(M_S^{\mu}M_S^{\mu\dagger}\big)+|\tr_S M_S^{\mu}|^2}{m(m+1)},
\end{equation}
which yields Eq.~\eqref{eq:Feff_unprot}. Finally, the success-weighted average conditional fidelity is the ratio
\begin{align}
    \bar F_{\mathrm{cond,w}}^{(\mathrm{unprot})}(t,s)
    &=
    \frac{\bar F_{\mathrm{eff}}^{(\mathrm{unprot})}(t,s)}{\bar p_{\mathrm{succ}}^{(\mathrm{unprot})}(t,s)}
    \nonumber\\
    &=
    \frac{\operatorname{Tr}_S\!\big(M_S^{\mu}(s)M_S^{\mu\dagger}(s)\big)+\bigl|\operatorname{Tr}_S M_S^{\mu}(s)\bigr|^{2}}{(m+1)\,\operatorname{Tr}_S G_S^{\mu}(s)},
\end{align}
which is Eq.~\eqref{eq:Fcond_unprot}.

\subsection*{Specialization underlying the timing proposition}

For the coherent $V$ sender, $\Lambda$ receiver, and $S=\{0,2\}$ used in Proposition~\ref{prop:fixed_timing}, set $\Delta=T-\tau$ and $z=e^{-\Gamma\Delta}$. Direct restriction of Eqs.~\eqref{eq:KV_explicit} and~\eqref{eq:KLambda_explicit} gives
\begin{align}
[c_A^{V,02}(\tau)]^2
&=\frac{1+e^{-2\Gamma\tau}}{2},
&
[c_C^\Lambda(T)]^2
&=e^{-2\Gamma T},
\label{eq:app_timing_c_factors}\\
\tr_S G_{02}^{V}(\Delta)
&=\frac{3+z^2}{2},
\label{eq:app_timing_traces_1}\\
\tr_S(M_{02}^{V}M_{02}^{V\dagger})
&=\frac{z^2+2z+5}{4},
\label{eq:app_timing_traces_2}\\
\bigl|\tr_S M_{02}^{V}(\Delta)\bigr|^2
&=\frac{z^2+6z+9}{4}.
\label{eq:app_timing_traces_3}
\end{align}
Substitution into Eqs.~\eqref{eq:psucc_unprot} and~\eqref{eq:Fcond_unprot}, with $m=2$, yields
\begin{align}
\frac{\bar p_{\mathrm{fix}}(\tau;T)}{p_C(T)}
&=\frac{(1+e^{-2\Gamma\tau})
[3+e^{-2\Gamma(T-\tau)}]}{8},
\\
\bar F_{\mathrm{cond,w}}^{(\mathrm{fix})}(\tau;T)
&=\frac{z^2+4z+7}{3(z^2+3)},
\end{align}
which are Eqs.~\eqref{eq:fixed_timing_success} and~\eqref{eq:fixed_timing_fidelity}. In the active-constraint range $F_0>F_{\mathrm{nr}}(x)>7/9$, the fidelity boundary follows from
\begin{equation}
(3F_0-1)z^2-4z+(9F_0-7)=0.
\end{equation}
The root in the physical interval $e^{-\Gamma T}\leq z\leq1$ is
\begin{equation}
\begin{aligned}
z_F
&=\frac{2-\sqrt{4-(3F_0-1)(9F_0-7)}}{3F_0-1}\\
&=\frac{2-\sqrt{3(1-F_0)(9F_0-1)}}{3F_0-1},
\end{aligned}
\end{equation}
and $\Gamma\tau_F=\Gamma T+\ln z_F$, as used in Eq.~\eqref{eq:earliest_feasible_time}.

\section{Full qutrit teleportation protocol and Bell measurement definitions}\label{app:qutrit_teleportation}

This appendix gathers the explicit qutrit-level definitions of the Bell basis and the unitary correction. We adopt one fixed convention and show the contraction so the correction is unambiguous.

We work with three qutrits $A$, $B$, $C$. Alice holds the unknown encoded input $A$ and resource half $B$, while Charlie holds $C$; the resource pair $BC$ is initialized in $|\Phi_3\rangle_{BC}$ of Eq.~\eqref{eq:phi3_resource}. The qutrit Bell basis on $A\otimes B$ is generated by the discrete Heisenberg--Weyl shift and clock operators
\begin{equation}\label{eq:hw_operators}
    X|j\rangle=|j\oplus 1\rangle,
    \qquad
    Z|j\rangle=\omega^{j}|j\rangle,
    \qquad
    \omega=e^{2\pi i/3},
\end{equation}
where $\oplus$ denotes addition modulo $3$. We reserve $m$ for the encoded-support dimension and choose throughout the convention in which $X^rZ^n$ acts on the $B$ side,
\begin{equation}\label{eq:qutrit_bell_basis}
    |\Phi_{r,n}\rangle_{AB}
    =
    \big(I_A\otimes X_B^{r}Z_B^{n}\big)|\Phi_3\rangle_{AB},
    \qquad
    r,n\in\{0,1,2\}.
\end{equation}
The nine vectors $|\Phi_{r,n}\rangle_{AB}$ form an orthonormal basis of $\mathcal{H}_A\otimes\mathcal{H}_B$.

For the input $|\psi_S\rangle_A=\sum_j a_j|j\rangle_A$ supported on $S\subseteq\{0,1,2\}$, the joint state on $ABC$ before the Bell measurement is
\begin{equation}\label{eq:appC_full_state}
    |\psi_S\rangle_A\otimes|\Phi_3\rangle_{BC}
    =
    \frac{1}{\sqrt{3}}\sum_{j,l} a_j |j\rangle_A|l\rangle_B|l\rangle_C.
\end{equation}
A direct contraction with the Bell bra $\langle\Phi_{r,n}|_{AB}=(1/\sqrt{3})\sum_k\omega^{-nk}\langle k|_A\langle k\oplus r|_B$ gives, conditional on outcome $(r,n)$,
\begin{align}\label{eq:appC_contraction}
    {}_{AB}\langle\Phi_{r,n}|\big(|\psi_S\rangle_A\otimes|\Phi_3\rangle_{BC}\big)
    &=
    \frac{1}{3}\sum_{k}\omega^{-nk}a_k|k\oplus r\rangle_C
    \nonumber\\
    &=
    \frac{1}{3}\,X_C^{r}Z_C^{-n}|\psi_S\rangle_C,
\end{align}
where the second equality uses
\begin{align}\label{eq:appC_XZ_action}
    X_C^r Z_C^{-n}|\psi_S\rangle_C
    &=
    \sum_j \omega^{-nj}a_j|j\oplus r\rangle_C
    \nonumber\\
    &=
    \sum_k\omega^{-nk}a_k|k\oplus r\rangle_C.
\end{align}
Charlie's qutrit is therefore left in the state $X_C^{r}Z_C^{-n}|\psi_S\rangle_C$ up to the global $1/3$ amplitude, and Charlie applies the Heisenberg--Weyl correction
\begin{equation}\label{eq:qutrit_correction}
    U_{r,n}^{(C)}=Z_C^{n}X_C^{-r},
\end{equation}
which is the inverse of the operator that acted on Charlie's state in Eq.~\eqref{eq:appC_contraction} (and equivalently the Hermitian conjugate of $X_C^{r}Z_C^{-n}$). The Bell-basis convention and correction $U_{r,n}^{(C)}$ apply to all nine Bell outcomes $(r,n)\in\{0,1,2\}^{2}$, each occurring with probability $1/9$ in the noiseless case. Averaging over the uniform distribution of Bell outcomes recovers the standard noiseless teleportation channel.

For the noisy protocol, write the encoded input as $Q_S|\varphi\rangle$ and let $D_A$ denote the sender map remaining immediately before the Bell measurement. If $C$ has evolved through $K_C$, the same contraction gives the complete transfer branch
\begin{align}
&{}_{AB}\langle\Phi_{r,n}|
\left[(D_AQ_S|\varphi\rangle)_A\otimes
(I_B\otimes K_C)|\Phi_3\rangle_{BC}\right]
\nonumber\\
&\hspace{7em}=
\frac{1}{3}K_CX_C^rZ_C^{-n}D_AQ_S|\varphi\rangle.
\label{eq:appC_noisy_contraction}
\end{align}
The final column is interpreted in $C$ through the fixed computational-basis
identification between $A$ and $C$; the ideal logical target is
$Q_S^{(C)}|\varphi\rangle$.
After a maximal exact sender PR at $\tau$, the common sender amplitude is $c_A(\tau)$ and $D_A=K_A(T-\tau)$. Thus Eq.~\eqref{eq:appC_noisy_contraction} is exactly $(c_A/3)\mathcal T_{r,n}|\varphi\rangle$, with $\mathcal T_{r,n}$ defined in Eq.~\eqref{eq:full_branch_transfer}.

The role of the Bell factor differs between the two receiver architectures. In Architecture A, $M_{C,\mathrm{EA}}K_C=c_CI$ and the unitary $U_{r,n}^{(C)}$ removes $X_C^rZ_C^{-n}$. Each outcome then gives the same contribution with squared amplitude $1/9$; accepting and summing all nine outcomes cancels that factor, which is why Eqs.~\eqref{eq:psucc_unprot}--\eqref{eq:Fcond_unprot} contain no explicit $1/9$. In Architecture B, by contrast, the optimal nonunitary scale depends on $(r,n)$. The per-outcome exact-recovery probability is
\begin{equation}
p_{r,n}=\frac{c_A^2(\tau)}{9}
\sigma_{\min}^2[\mathcal T_{r,n}(\tau;T)],
\end{equation}
so the factor $1/9$ must remain in the outcome-summed expression, giving Eq.~\eqref{eq:full_branch_probability}. The branch-dependent filter already maps the output back through $Q_S^{(C)}$ and therefore includes the Bell correction.

\section{Local qutrit jump operators and supplementary sender-support diagnostics}\label{app:jump_operators}

This appendix collects the explicit matrix forms of the local single-qutrit jump-operator topologies used throughout the main text. Each topology specifies a single coherent Lindblad operator $L^\mu$ acting on $\mathbb{C}^3$ in the computational basis $\{|0\rangle,|1\rangle,|2\rangle\}$. The pathways displayed within a given $L^\mu$ are therefore summed coherently before $L^{\mu\dagger}L^\mu$ is formed; replacing them by independent jump operators would remove the associated cross terms and define a different model. We use the labels $\Lambda$, $V$, and ``all'' as representative jump-operator topologies that are local to a single qutrit; no energy ordering of $|0\rangle,|1\rangle,|2\rangle$ is assumed unless explicitly stated. Each topology can be installed on the sender qutrit $A$, the receiver qutrit $C$, or any other single qutrit in the protocol.

The $\Lambda$-type topology has the form
\begin{equation}\label{eq:appD_lambda}
    L^{\Lambda}=
    \begin{pmatrix}
        0 & 0 & 0\\
        1 & 0 & 0\\
        1 & 0 & 0
    \end{pmatrix}
    =|1\rangle\langle 0|+|2\rangle\langle 0|,
\end{equation}
describing jump pathways from level $|0\rangle$ to both $|1\rangle$ and $|2\rangle$. One has $L^{\Lambda\dagger}L^{\Lambda}=2|0\rangle\langle 0|$, which is diagonal in the computational basis.

The $V$-type topology is
\begin{equation}\label{eq:appD_V}
    L^{V}=
    \begin{pmatrix}
        0 & 0 & 0\\
        0 & 0 & 0\\
        1 & 1 & 0
    \end{pmatrix}
    =|2\rangle\langle 0|+|2\rangle\langle 1|,
\end{equation}
describing jump pathways from $|0\rangle$ and $|1\rangle$ into the common level $|2\rangle$. One has $L^{V\dagger}L^{V}=(|0\rangle+|1\rangle)(\langle 0|+\langle 1|)$, whose range is one-dimensional and not aligned with any single computational basis vector.

The all-transition topology is
\begin{equation}\label{eq:appD_all}
    L^{\mathrm{all}}=
    \begin{pmatrix}
        0 & 0 & 0\\
        1 & 0 & 0\\
        1 & 1 & 0
    \end{pmatrix},
\end{equation}
combining the $\Lambda$- and $V$-type pathways. Its bright subspace is two-dimensional, with two nonzero eigendirections, and $L^{\mathrm{all}\dagger}L^{\mathrm{all}}$ includes an off-diagonal coherence between $|0\rangle$ and $|1\rangle$ that is absent from the $\Lambda$-type case. Here ``all-transition'' is shorthand for the three directed transitions in the chosen orientation of the level graph $|0\rangle\to|1\rangle$, $|0\rangle\to|2\rangle$, $|1\rangle\to|2\rangle$, not all six bidirectional transitions between the three basis states; the same convention is used in Section~\ref{subsec:model_compare}.

The associated Markovian no-jump Kraus operator on a single qutrit is
\begin{equation}\label{eq:appD_K_single_qutrit}
\begin{aligned}
    K_{0}^{\mu}(t)
    &=\exp\!\left[\left(-iH_{\mathrm{sys}}
      -\frac{\Gamma}{2}L^{\mu\dagger}L^{\mu}\right)t\right],\\
    &\hspace{5em}\mu\in\{\Lambda,V,\mathrm{all}\}.
\end{aligned}
\end{equation}
For the zero-Hamiltonian numerical model used here, the bright eigenmodes of $L^{\mu\dagger}L^{\mu}$ determine the directions in $\mathbb{C}^3$ that are most strongly suppressed by the no-jump branch. At finite times, the maximal PR success probability strictly exceeds the full-space EA-WMR value exactly when the encoded support $\mathcal{H}_S$ has trivial intersection with the right-singular subspace of $K_0^{\mu}(t)$ associated with its smallest singular value, as characterized by Eq.~\eqref{eq:gain_ratio_exact}. With a Hamiltonian that does not commute with $L^{\mu\dagger}L^\mu$, the singular directions must instead be computed from $K_0^{\mu\dagger}(t)K_0^\mu(t)$.

\begin{figure*}[tbp]
    \centering
    \includegraphics[width=0.49\textwidth]{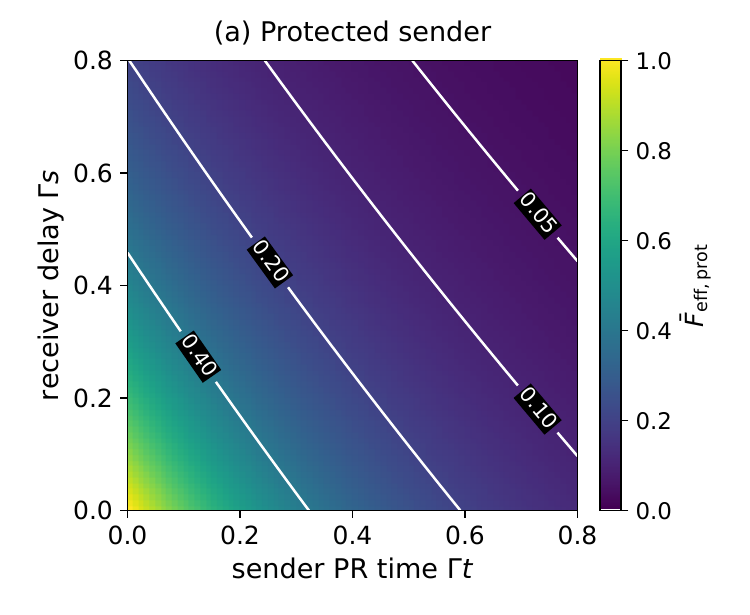}
    \hfill
    \includegraphics[width=0.49\textwidth]{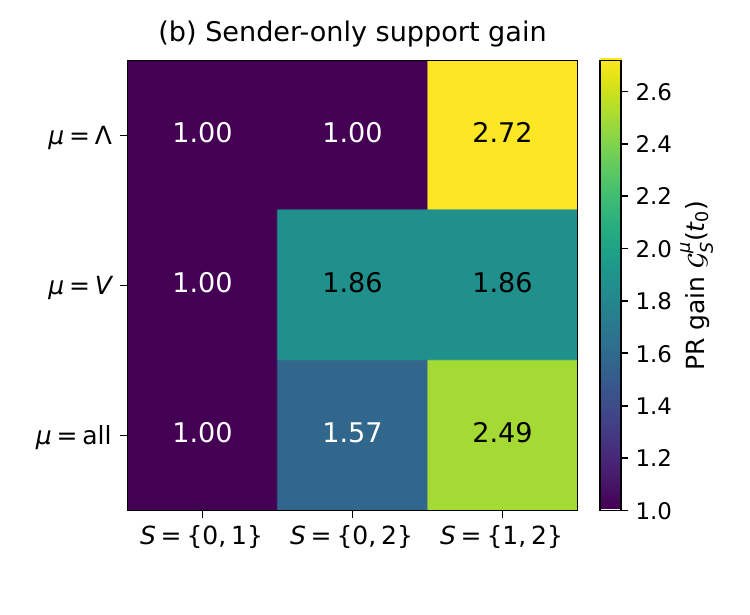}
    \caption{Supplementary recovery diagnostics. (a) Protected-sender
    Architecture-A benchmark,
    $\bar F_{\mathrm{eff,prot}}=p_{\mathrm{tot,max}}^{(\mathrm{prot})}$,
    for $\mu=V$, $\chi=\Lambda$, and $\Pi_{02}$ on
    $\Gamma t,\Gamma s\in[0,0.8]$; the labeled contours are
    $\{0.05,0.1,0.2,0.4\}$. This heralded success-weighted quantity
    equals the postselection probability because its conditional
    fidelity is one, and it is not a deterministic encoded-qubit
    fidelity. (b) Sender-side support gain at $\Gamma t_0=0.5$ across
    the three topologies and two-level supports. A value of one marks
    a nontrivial intersection with the full-space minimum right-singular subspace;
    a value above one is a strict sender-level PR gain. Panel (b) contains
    neither receiver dynamics nor a residual sender interval and is not
    a protocol-level comparison.}
    \label{fig:supp_sender_diagnostics}
\end{figure*}

A heatmap of the protected-sender effective fidelity $\bar F_{\mathrm{eff,prot}}(t,s)$ on the $(\Gamma t,\Gamma s)$ plane, for the representative configuration $\mu=V$, $\chi=\Lambda$, $S=\{0,2\}$, is reported as Fig.~\ref{fig:supp_sender_diagnostics}(a). The displayed quantity equals
$\bar F_{\mathrm{eff,prot}}(t,s)=p_{\mathrm{tot,max}}^{(\mathrm{prot})}(t,s)=[\sigma_{\min}^{S}(K_A^V(t))]^{2}\,\sigma_{\min}^{2}(K_C^\Lambda(t+s))$, a heralded success-weighted quantity equal to the total postselection probability because the conditional fidelity is unity on success in this benchmark; it should not be compared directly with the deterministic encoded-qubit teleportation reference $2/3$. The plot resolves the informative window $\Gamma t,\Gamma s\in[0,0.8]$ and superimposes the contour levels $\{0.05,0.1,0.2,0.4\}$ on the color map. The receiver-side full-space factor over the total time $T=t+s$ is the dominant bottleneck within the stipulated outcome-independent receiver architecture.

As a compact summary across topologies and supports, Fig.~\ref{fig:supp_sender_diagnostics}(b) reports the support-restricted PR gain
$\mathcal{G}_S^\mu(t_0)=[\sigma_{\min}^{S}(K_A^\mu(t_0))]^{2}/\sigma_{\min}^2(K_A^\mu(t_0))$
at a fixed evaluation time $\Gamma t_0=0.5$ as a $3\times 3$ heatmap with rows indexed by the sender topology $\mu\in\{\Lambda,V,\mathrm{all}\}$ and columns by the encoded support $S\in\{01,02,12\}$. The numerical value is overlaid on each cell, so the heatmap doubles as a table of gain factors.

\section{Whole-Interval OU Note and the Need for a Process Tensor under Intermediate Intervention}\label{app:OU_process_tensor}

\begin{figure*}[tbp]
    \centering
    \includegraphics[width=0.9\textwidth]{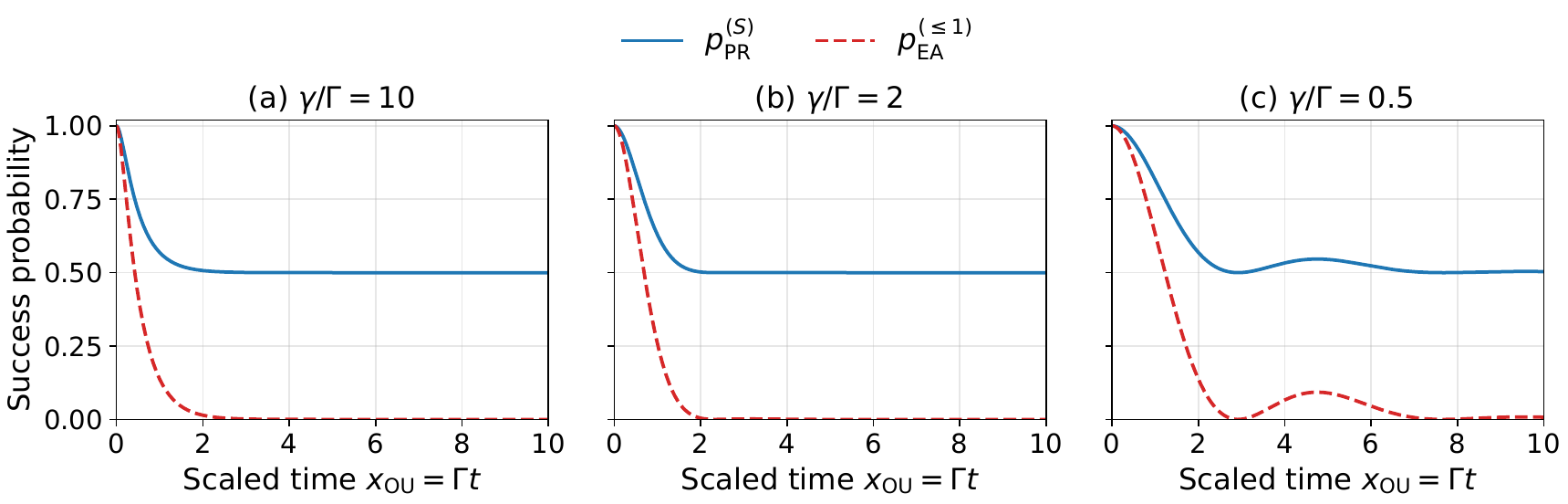}
    \caption{Supporting whole-interval mechanism (not a qutrit teleportation protocol-level result). The plotted quantities are the support-restricted probability $p_{\mathrm{PR}}^{(S)}$ (solid) and the zero/single-excitation-sector benchmark $p_{\mathrm{EA}}^{(\leq1)}$ (dashed) of Eq.~\eqref{eq:two_qubit_pr_vs_ea}, versus $x_{\mathrm{OU}}=\Gamma t$. Panels (a)--(c): $\gamma/\Gamma=10,2,0.5$. Memory-induced revivals in $|G(t)|^2$ lift both curves; the persistent solid--dashed gap reflects the support-restricted singular-value mechanism. At zeros of $G(t)$, the dashed value is the continuous singular-value scale and exact sector-wide EA-WMR is undefined. The dashed curve is not the exact full-space EA-WMR benchmark unless $|G_2(t)|\geq|G(t)|$.}
    \label{fig:ou_twoqubit_compare}
\end{figure*}

The Ornstein--Uhlenbeck no-jump amplitude $G(t)$ derived in Section~\ref{subsec:two_qubit_supporting} describes a single uninterrupted dissipative interval $[0,t]$ with the system--bath state initialized at the start of that interval. The closed-form expression in Eq.~\eqref{eq:G_solution} therefore applies to scenarios in which the PR (or EA-WMR) operation is implemented at the end of the noisy interval, after which the bath history is allowed to terminate. This is the configuration analyzed in the main text.

By contrast, scenarios in which an intermediate operation is inserted at some time $0<t_1<t$, after which the system continues to interact with the same correlated bath, do \emph{not} admit a semigroup composition of the form
\begin{equation}\label{eq:OU_no_semigroup}
    K^{\mathrm{OU}}(t)\stackrel{?}{=}K^{\mathrm{OU}}(t-t_1)\,K^{\mathrm{OU}}(t_1),
\end{equation}
because the bath retains memory of the system trajectory across the intermediate intervention. A consistent description must retain these multi-time system--bath correlations; one systematic option is the process-tensor formalism, in which the correlations are encoded in a multilinear map and intermediate operations are inserted at its appropriate ``slots''~\cite{pollock2018nonmarkovian}. We do not pursue that analysis here. Throughout the main text we therefore restrict the OU non-Markovian results to whole-interval configurations and reserve the analysis of intermediate intervention in correlated baths for future work.

Figure~\ref{fig:ou_twoqubit_compare} compares the support-restricted PR probability with the zero/single-excitation-sector EA-WMR benchmark of Eq.~\eqref{eq:two_qubit_pr_vs_ea} for $\gamma/\Gamma\in\{10,2,0.5\}$. The full four-dimensional benchmark also requires $G_2(t)$, as in Eq.~\eqref{eq:two_qubit_full_ea}. These whole-interval curves illustrate the singular-value mechanism; they are not a qutrit teleportation protocol-level result.

\section{Optional block-level extension not used in the main numerical examples}\label{app:block_extension}

The local-noise model used in the main text places independent Markovian baths on Alice's qutrit $A$ and Charlie's receiver qutrit $C$, with the entangled partner $B$ taken as ideal. The same singular-value argument extends without change to a joint sender block $AB$ that shares a collective bath. We summarize this block-level extension here for completeness; it is not used by any of the figures or numerical conclusions of the main text.

Let the sender block $AB$ have collective Hamiltonian and jump operator
\begin{equation}\label{eq:appBlock_HL}
    H_{AB}=\omega(J_z\otimes\id+\id\otimes J_z),
    \qquad
    L_{AB}=J_-\otimes\id+\id\otimes J_-,
\end{equation}
with the qutrit ladder generators of the standard $\mathfrak{su}(2)$ representation. The associated joint no-jump propagator is
\begin{equation}\label{eq:appBlock_K_AB}
    K_{AB,0}(t)=\exp\!\left[\bigl(-iH_{AB}-\tfrac{\Gamma_{AB}}{2}L_{AB}^\dagger L_{AB}\bigr)t\right],
\end{equation}
acting on $\mathcal{H}_{AB}\simeq\mathbb{C}^3\otimes\mathbb{C}^3$. Let $\Pi_{AB}$ denote the projector onto the encoded sender subspace within $\mathcal{H}_{AB}$, with rank $m_{AB}$, and let $Q_{AB}:\mathbb{C}^{m_{AB}}\to\mathcal{H}_{AB}$ be an isometry whose columns form an orthonormal basis of that subspace, $Q_{AB}^\dagger Q_{AB}=I_{m_{AB}}$ and $Q_{AB}Q_{AB}^\dagger=\Pi_{AB}$. Define the restricted block matrix
\begin{equation}\label{eq:appBlock_B}
    B_{AB}(t)=K_{AB,0}(t)\,Q_{AB}\in\mathcal{L}(\mathbb{C}^{m_{AB}},\mathcal{H}_{AB}).
\end{equation}
The successful PR operator on the joint block is the Moore--Penrose pseudoinverse of the restricted block map composed with $Q_{AB}$,
\begin{align}\label{eq:appBlock_R_PR}
    R_{AB,\mathrm{PR}}(t)
    &=
    Q_{AB}\,B_{AB}^{+}(t)
    \nonumber\\
    &=
    Q_{AB}\bigl[B_{AB}^\dagger(t)B_{AB}(t)\bigr]^{-1}B_{AB}^\dagger(t),
\end{align}
in direct analogy with the single-qutrit construction of Section~\ref{sec.3}, and the corresponding successful weak-measurement element is $M_{AB,\mathrm{PR}}(t)=c_{AB}(t)R_{AB,\mathrm{PR}}(t)$ with $c_{AB}(t)=\sigma_{\min}(B_{AB}(t))$. The maximal sender-side success factor is therefore
\begin{equation}\label{eq:appBlock_pPR_max}
    p_{AB,\mathrm{PR}}^{\max}(t)=\sigma_{\min}^{2}\!\bigl(B_{AB}(t)\bigr),
\end{equation}
and the protected total success probability factorizes against the receiver-side full-space EA-WMR factor:
\begin{equation}\label{eq:appBlock_p_max}
    p_{\mathrm{tot,max}}^{(\mathrm{prot,AB})}(t,s)
    =
    \sigma_{\min}^2\!\bigl(B_{AB}(t)\bigr)\,
    \sigma_{\min}^2\!\bigl(K_{C,0}(t+s)\bigr).
\end{equation}
Equation~\eqref{eq:appBlock_p_max} is the joint-block analog of Eq.~\eqref{eq:async_total_probability_optimal} in the main text. The unprotected counterpart is obtained by composing $K_{AB,0}(s)M_{AB,\mathrm{PR}}(t)K_{AB,0}(t)$ on the joint block and reading off the residual block dynamics, in direct parallel with the unprotected-sender Architecture-A variant of Section~\ref{subsec:architecture_A_unprotected}.

\bibliography{ref}

\end{document}